\documentclass[%
aip,
 amsmath,amssymb,
 reprint,%
]{revtex4-1}

\usepackage{graphicx}
\usepackage{dcolumn}
\usepackage{bm}

\usepackage{mathptmx}
\usepackage{etoolbox}
\usepackage{enumitem}

\usepackage{hyperref}
\usepackage{color}
\usepackage{cancel}

\newcommand{\LC}{\ell c}
\newcommand{\nc}{\text{nc}}
\newcommand{\opt}{\text{opt}}
\newcommand{\EL}{\textsuperscript{EL}}
\newcommand{\sctf}{s_f}
\newcommand{\trel}{t_{\text{R}}}

\DeclareMathOperator{\sgn}{sgn}

\makeatletter
\def\@email#1#2{%
 \endgroup
 \patchcmd{\titleblock@produce}
  {\frontmatter@RRAPformat}
  {\frontmatter@RRAPformat{\produce@RRAP{*#1\href{mailto:#2}{#2}}}\frontmatter@RRAPformat}
  {}{}
}%
\makeatother

\begin{document}


\title{Optimal synchronisation to a limit cycle} 

\author{C.~Ríos-Monje}
 \affiliation{Física Teórica, Universidad de Sevilla, Apartado de
  Correos 1065, E-41080 Sevilla, Spain}
\author{C.~A.~Plata}
 \affiliation{Física Teórica, Universidad de Sevilla, Apartado de
  Correos 1065, E-41080 Sevilla, Spain}
\author{D.~Guéry-Odelin}
\affiliation{Laboratoire Collisions, Agr\'egats, R\'eactivit\'e, FeRMI, Université  Toulouse III - Paul Sabatier, 118 Route de Narbonne, 31062 Toulouse Cedex 09, France}
\affiliation{Institut Universitaire de France}
\author{A.~Prados} 
\email{prados@us.es}
 \affiliation{Física Teórica, Universidad de Sevilla, Apartado de
  Correos 1065, E-41080 Sevilla, Spain}
  
\date{\today}

\begin{abstract}
In the absence of external forcing, all trajectories on the phase plane of the van der Pol oscillator tend to a closed, periodic, trajectory---the limit cycle---after infinite time. Here, we drive the van der Pol oscillator with an external time-dependent force to reach the limit cycle in a given finite time. Specifically, we are interested in minimising the non-conservative contribution to the work when driving the system from a given initial point on the phase plane to any final point belonging to the limit cycle. There appears a speed limit inequality, which expresses a trade-off between the connection time and cost---in terms of the non-conservative work. We show how the above results can be { generalized to the  broader family of non-linear oscillators given by} the Liénard equation. Finally, we also look into the problem of minimising the total work done by the external force.
\end{abstract}

\maketitle

\begin{quotation}
Self-sustained oscillations are relevant in physics---e.g. in electronics, lasers, or active matter, but also in many other contexts: neuroscience, physiology, economics, to name just a few.\cite{ginoux_van_2012,jenkins_self-oscillation_2013} A key feature of self-oscillatory systems is the existence of a stable limit cycle, which appears as a consequence of the non-linearity of the damping force. The limit cycle is stable because neighbouring trajectories approach it in the long-time limit, i.e. after a certain typical relaxation time $t_R$. Here, importing ideas from the field of shortcuts to adiabaticity in quantum mechanics\cite{guery-odelin_shortcuts_2019} and swift state-to-state transformations (SST) in classical and stochastic systems,\cite{guery-odelin_driving_2023} we address the problem of shortcutting the relaxation to the limit cycle by driving the system with a suitable external force during a given time $t_f<t_R$. Specifically, we aim at building a SST that minimises the non-conservative work. The emergence of a speed limit inequality is shown, with a trade-off between the connection time and the non-conservative work. This inequality entails that, for small damping, when the natural relaxation time of the system to the limit cycle is very long, the relaxation to the limit cycle can be accelerated by a very large factor while keeping the energetic cost finite. Interestingly, the developed framework for the van der Pol oscillator naturally extends to the more general case of the Liénard equation, with tiny changes. Also, the minimisation of the total work, including both the conservative and non-conservative contributions, is investigated. 
\end{quotation}

\section{Introduction}\label{sec:intro}

Stable limit cycles appear in systems that show oscillatory behaviour in the long-time limit without any time-periodic driving. In other words, limit cycles correspond to self-sustained oscillations, which are present in a wide variety of { non-linear} systems. In physics, self-sustained oscillations in non-linear circuits were the motivation for van der Pol's pioneering work\cite{van_der_pol_lxxxviii_1926}, see  Ref.~\onlinecite{ginoux_van_2012} for a review. More currently, they are being investigated in the field of active matter.\cite{sakaguchi_limit-cycle_2009,alicki_leaking_2021,kotwal_active_2021,liu_viscoelastic_2021,tucci_modeling_2022} Also, self-oscillations are  often found in biological systems, for example they have been shown to be  relevant for circadian rythms\cite{leloup_limit_1999,roenneberg_modelling_2008} or the migration of cancer cells in confined environments.~\cite{bruckner_stochastic_2019}

The van der Pol equation is a paradigmatic model for self-sustained oscillations. In dimensionless variables, it reads
\begin{equation}\label{eq:van-der-Pol}
    \ddot{x}+\mu (x^2-1)\dot{x}+x=0,
\end{equation}
where $\mu>0$ is a parameter, to which we will refer as the damping constant. Variations of the van der Pol equation have been employed in many fields, e.g. to electronically simulate nerve axons,\cite{fitzhugh_impulses_1961,nagumo_active_1962} to understand the dynamics of elastic excitable media,\cite{cartwright_dynamics_1999} or to model self-sustained oscillations in active matter.\cite{romanczuk_active_2012} The undriven van der Pol oscillator takes an infinite time to reach the limit cycle, with a relaxation time that decreases with the damping constant $\mu$. For small damping $\mu\ll 1$, the natural relaxation time $t_R$ to the limit cycle is very long, $t_R=O(\mu^{-1})$.\cite{strogatz_nonlinear_2024}

Only very recently,\cite{impens_shortcut_2023} the problem of accelerating the relaxation to the limit cycle has been addressed, in the context of SST recently introduced in classical and stochastic systems.\cite{guery-odelin_driving_2023} The general idea of SST, which is rooted in the field of quantum shortcuts to adiabaticity,\cite{guery-odelin_shortcuts_2019} is to introduce a suitable driving to make the system reach the desired target state in a time shorter than the natural relaxation time. Still, neither the energetic cost of such an acceleration nor the physical implications thereof has been investigated for the synchronisation to a limit cycle.

In this work, we aim at investigating the energetic cost of driving the van der Pol oscillator to its limit cycle in a finite time. More specifically, we are interested in engineering an optimal driving $F(t)$---to be added to the right hand side (rhs) of Eq.~\eqref{eq:van-der-Pol}---that minimises the dissipative work done by the non-conservative { non-linear} force $F_{\nc}(x,\dot{x})\equiv -\mu(x^2-1)\dot{x}$. This problem has strong similarities with the minimisation of the irreversible work in stochastic systems,\cite{schmiedl_optimal_2007,aurell_optimal_2011,plata_optimal_2019,zhang_work_2020,guery-odelin_driving_2023} although the van der Pol oscillator is not coupled to a heat bath. Also, it raises the question of the trade-off between operation time and cost, as measured by the non-conservative work, and the possible emergence of speed limit inequalities.\cite{sivak_thermodynamic_2012,deffner_quantum_2017,okuyama_quantum_2018,shiraishi_speed_2018,shanahan_quantum_2018,funo_speed_2019,shiraishi_speed_2020,plata_finite-time_2020,deffner_quantum_2020,ito_stochastic_2020,van_vu_geometrical_2021,prados_optimizing_2021,lee_speed_2022,patron_thermal_2022,dechant_minimum_2022,guery-odelin_driving_2023}

The minimisation of the non-conservative work is carried out by considering that the system starts from a given point on the phase plane and reaches any point of the limit cycle in a given connection time $t_f$---ideally, much shorter than the natural relaxation time $t_R$. When the force is not bounded, which is the case we address in this paper, this minimisation problem can be tackled with the tools of variational calculus, incorporating the variable endpoint via the so-called transversality condition.\cite{gelfand_calculus_2000,liberzon_calculus_2012} Remarkably, our analysis show that many of the results, including the optimal endpoint over the limit cycle, can be obtained without having to know the explicit expression for the limit cycle.

In this work, we also consider possible generalisations of the above problem. First,  we will show that practically all the results found for the minimisation of the non-conservative work in the van der Pol case extend, with tiny changes, to the Liénard equation---which is a model that covers a much broader family of oscillators, including van der Pol's as a particular case.  Second, we will tackle the minimisation of the total work done by the external force, including both the conservative and non-conservative contributions. 

The structure of the paper is as follows. Section~\ref{sec:model-basic-eqs} presents the model and some basic equations that we need for our analysis. In Sec.~\ref{sec:minimisation-nc-work-unbounded}, we carry out the minimisation of the non-conservative work. We derive the Euler-Lagrange equations for the optimal path in  phase space in Sec.~\ref{sec:EL-equation}. Therefrom, we obtain explicit expressions for the optimal path and force in Sec.~\ref{sec:optimal-x-and-F}. The transversality condition for the variable endpoint problem is analysed in Sec.~\ref{sec:transvers-cond-var-endpoint}. Using the results of the previous sections, the explicit expression for the minimum non-conservative work is derived in Sec.~\ref{sec:min-non-conserv-work}. We illustrate our results and discuss their physical implications in Sec.~\ref{sec:phys-interpret-1}. Section~\ref{sec:general-results} is devoted to generalising our results to more complex scenarios: the Liénard equation in Sec.~\ref{sec:lienard-eq} and the minimisation of the total work in Sec.~\ref{sec:min-total-work}. Finally, we present the main conclusions of our work and some perspectives for further research in Sec.~\ref{sec:conclusions}.

\section{Model}\label{sec:model-basic-eqs}

As anticipated in the introduction, we are interested in investigating the van der Pol oscillator driven by an external force $F(t)$, 
\begin{equation}
    \ddot{x} + \mu (x^2-1) \dot{x} + x = F(t),
    \label{eq:driven-vdP}
\end{equation}
to accelerate the relaxation towards the limit cycle, which is described by an implicit function of $(x,\dot{x})$ that also depends on the parameter $\mu$, $\chi_{\LC}(x,\dot{x};\mu)=0$. In the following, we refer to Eq.~\eqref{eq:driven-vdP} as the driven van der Pol equation (dvdPE). For our purposes, it is convenient to explicitly introduce the phase plane point as 
\begin{equation}
    \bm{x}(t)\equiv (x_1(t),x_2(t)), \qquad x_1 \equiv x, \quad x_2\equiv \dot{x},
\end{equation} 
and rewrite the dvdPE \eqref{eq:driven-vdP} as a system of first order differential equations
\begin{equation}
    \dot{\bm{x}} = \begin{pmatrix}
        x_2 \\ -\mu(x_1^2-1)x_2-x_1+F
    \end{pmatrix}
    \equiv  \bm{f}(\bm{x};F),
    \label{eq:dvdPE-phase-space}
\end{equation}
i.e.
\begin{subequations}\label{eq:f1-f2-def}
\begin{align}
    \dot{x}_1&=f_1(x_2)\equiv x_2, \\
    \dot{x}_2&=f_2(x_1,x_2;F)\equiv -\mu(x_1^2-1)x_2-x_1+F.
\end{align} 
\end{subequations}
Note that, in order to simplify our notation, we have omitted the time dependence both in the phase-space variables $(x_1(t),x_2(t))$ and in the force $F(t)$.

We would like to reach the limit cycle by applying the driving force $F(t)$ for a given finite time $t_f$. That is, we want to drive the system from a given initial state $(x_{10},x_{20})$, to a final endpoint of the limit cycle, i.e. $(x_{1f},x_{2f})$ with $\chi_{\LC}(x_{1f},x_{2f};\mu)=0$, in a finite time $t_f$. Once the limit cycle is reached at $t=t_f$, the force is switched off, i.e. $F(t)=0$ for $t\ge t_f$, so that the system remains over the limit cycle $\forall t\ge t_f$. 

We aim at optimising the energetic cost of the driving described above. Bringing to bear the dvdPE \eqref{eq:driven-vdP} (or \eqref{eq:dvdPE-phase-space}), the work done on the system by the external force $F(t)$ can be split into a conservative part $\Delta E=E_f-E_0$, which only depends on the initial and final points on the phase plane, and a non-conservative part $W_{\nc}$, which depends on the whole trajectory of the system:
\begin{equation}
    W \equiv \int_0^{t_f} dt \, F(t) \, \dot{x}(t)=\Delta E+W_{\nc},
    \label{eq:W-def}
\end{equation}
with
\begin{equation}
    E(\bm{x})\equiv \frac{1}{2}\left(x_1^2+x_2^2\right), \quad W_{\nc}[\bm{x}]\equiv \mu \int_0^{t_f} dt \, (x_1^2-1)\, x_2^2.
    \label{E-and-Wnc-defs}
\end{equation}
Although the oscillator is not explicitly coupled to a heat bath, if we interpret $E$ as the internal energy of the oscillator, $W_{\nc}$ would be the heat dissipated to the environment due to the non-linear friction force.\footnote{Note that, at variance with simpler problems with linear damping, the sign of $W_{\nc}$ is not fixed because the factor $x_1^2-1$ changes sign at $x_1=1$.} In the following, we will mainly be interested in the minimisation of the non-conservative contribution to the work. This minimisation has to be done with the boundary conditions
\begin{subequations}\label{eq:bc}
  \begin{equation}\label{eq:ini-fin-fixed-points}
x_1(0)=x_{10}, \; x_2(0)={x}_{20}, \; x_1(t_f)=x_{1f}, \; x_2(t_f)={x}_{2f},
\end{equation}
\begin{equation}\label{eq:fLC-phase-space}
    \chi_{\LC}(\bm{x}_f;\mu)=0;
\end{equation}  
\end{subequations}
the latter condition enforces that $\bm{x}_f$ belongs to the limit cycle. In the following, it will be useful to introduce the maximum amplitude $x_{\LC}^{\max}$ of the oscillation over the limit cycle.

We remark the formal similarity of our minimisation problem with the minimisation of the irreversible work in systems with stochastic dynamics.\cite{aurell_optimal_2011,plata_optimal_2019,zhang_work_2020,guery-odelin_driving_2023,blaber_optimal_2023} Therein, the total work may be split into two contributions: (i) the free energy difference $\Delta F$, which is given by the initial and final equilibrium points (the analogous role here is played by $\Delta E$), and (ii) the irreversible work, which is a functional of the trajectory and it is positive definite (the analogous role here is played by $W_{\nc}$). Still, two key differences should be remarked: here, (i) the sign of $W_{\nc}$ is not necessarily positive, due to the change of sign of the non-linearity at $|x_1|=1$, which in turn is responsible for the emergence of a limit cycle, and (ii) the final point is not fixed, since we would like to join the initial point to any point of the limit cycle.\footnote{For fixed initial and final points of the trajectory on phase plane, $\Delta E$ has a given value. In that case, the problem of minimising the non-conservative work $W_{\nc}$ and the total work done by the external force $F(t)$ are always equivalent. See also Sec.~\ref{sec:min-total-work}.}

For an arbitrary value of the damping coefficient $\mu$, there is neither a closed analytical expression for the function $\chi_{\LC}(x_1,x_2;\mu)=0$ defining implicitly the limit cycle nor a closed analytical expression for the relaxation of the system towards it. Yet, in the small damping limit $\mu\ll 1$, a multiple scale analysis of Eq.~\eqref{eq:van-der-Pol} gives asymptotically valid expressions for both $\chi_{\LC}$ and the time evolution $\bm{x}(t)$ towards the long-time behaviour---e.g. see Ref.~\onlinecite{strogatz_nonlinear_2024}. Specifically, one has
\begin{subequations}\label{eq:vdP-sol-small-mu}
\begin{align}
\chi_{\LC} &\sim x_1^2+x_2^2-4+O(\mu), \label{eq:vdP-fLC-small-mu}\\
x_1(t) & \sim 2\left({1-\frac{r_0^2-
        4}{r_0^2}e^{-\mu t}}\right)^{\!\!\!\!\!-1/2} \!\!\!\!\!\cos{(t+\phi_0)} + O(\mu), \label{eq:vdP-x(t)-small-mu}\\
x_2(t) & \sim - 2\left({1-\frac{r_0^2-
        4}{r_0^2}e^{-\mu t}}\right)^{\!\!\!\!\!-1/2} \!\!\!\!\!\sin{(t+\phi_0)} + O(\mu), \label{eq:vdP-xdot(t)-small-mu}      
\end{align}
\end{subequations}
where
\begin{equation}
    r_0 = \sqrt{x_{10}^2+x_{20}^2}, \quad \phi_0 = \arctan{\frac{x_{20}}{x_{10}}}.
\end{equation}
For $\mu\ll 1$, the maximum amplitude of the limit cycle approaches $2$, 
\begin{equation}\label{eq:xlc-max-0}
   \widetilde{x}_{\LC}^{\;\max}\equiv \lim_{\mu\to 0^+} x_{\LC}^{\max}=2.
\end{equation}
The value of $\widetilde{x}_{\LC}^{\;\max}$ stems from the multiple scale expression in Eq.~\eqref{eq:vdP-x(t)-small-mu}. Also, a physical argument could be given as follows: for $\mu=0$, circular orbits with $x_1=A\cos (t+\phi_0)$, $x_2=-A\sin(t+\phi_0)$, for any $A$ are possible. For $\mu\ll 1$, the only value of $A$ that survives is that such the non-conservative work in Eq.~\eqref{E-and-Wnc-defs} vanishes over the circumference: this gives $A=2$. { On the one hand, it has been shown that $x_{\LC}^{\max}$ has a very weak dependence on $\mu$, being very close to $\widetilde{x}_{\LC}^{\;\max}=2$ for all $\mu$; namely $2\le x_{\LC}^{\max}\le 2.0672$.~\cite{turner_maximum_2015} On the other hand, the shape of the limit cycle strongly deviates from a circumference as $\mu$ increases.\cite{strogatz_nonlinear_2024}
}

\begin{figure}
    \centering
    \includegraphics[width=3in]{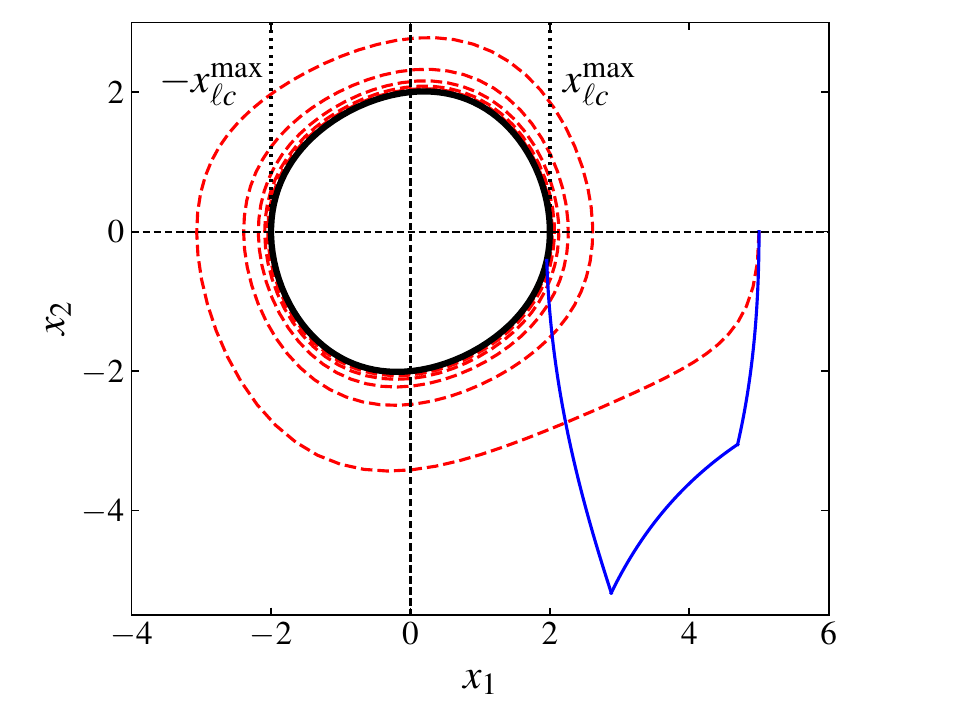}
    \caption{\label{fig:relax-to-limit-cycle}
    Illustration of a free trajectory and a driven trajectory on the phase plane $(x_1=x,x_2=\dot{x})$ of the van der Pol oscillator with $\mu=10^{-1}$. The thick line stands for the limit cycle of the system. { Over the limit cycle, $-x_{\LC}^{\max}\le x_1 \le x_{\LC}^{\max}$; note that the van der Pol equation is symmetrical under point reflection with respect to the origin, i.e. $(x_1,x_2)\to (-x_1,-x_2)$. Also plotted are the vertical lines $x_1=\pm x_{\LC}^{\max}$ (dotted). } The free trajectory (dashed red) reaches the limit cycle in an infinite time, whereas the driven trajectory (solid blue) reaches it in a finite time; here, $t_f=1$. The plotted data correspond to the numerical integration of the van der Pol equation. 
 }   
\end{figure}

For small damping, $\mu\ll 1$, the relaxation time of the system $\trel$ to the limit cycle is very long, of the order of $\mu^{-1}$. Specifically, we estimate $\trel=4\mu^{-1}$, for which $e^{-\mu\trel}\simeq 0.02$. In Fig.~\ref{fig:relax-to-limit-cycle}, we present a typical relaxation of the van der Pol oscillator to the limit cycle, for $\mu=0.1$, i.e. $\trel=40$. 

The small damping limit is especially relevant for building a SST to the limit cycle, by introducing a suitable driving $F(t)$, because of the long time scale of the natural relaxation. In the opposite limit of large damping, $\mu\gg 1$, the relaxation to the limit cycle is almost instantaneous in the undriven case, see e.g. Sec. 7.5 of Ref.~\onlinecite{strogatz_nonlinear_2024}. Therefore, in the following we mainly focus on the small damping limit $\mu\ll 1$.


\section{Minimisation of the non-conservative work}\label{sec:minimisation-nc-work-unbounded}

We would like to minimise $W_{\nc}$, which is a functional of the phase plane trajectory $\bm{x}$, as given by Eq.~\eqref{E-and-Wnc-defs}. We thus have a variational problem:
\begin{equation}
    W_{\nc}[\bm{x}]=\int_0^{t_f} dt\, L(\bm{x}), \quad L(\bm{x})=\mu (x_1^2-1) x_2^2,
\end{equation} 
where $L(\bm{x})=\mu (x_1^2-1) x_2^2$ is our ``Lagrangian''. This is a variational problem with constraints: $\dot{x}_1=f_1$, $\dot{x}_2=f_2$, as given by Eq.~\eqref{eq:f1-f2-def}. These dynamical constraints are incorporated to the minimisation problem by introducing time-dependent Lagrange multipliers. Moreover, the variable endpoint belonging to the limit cycle is included by considering the so-called transversality condition,\cite{gelfand_calculus_2000,liberzon_calculus_2012} { see also Appendix~\ref{app-transvers-cond}.}

\subsection{Euler-Lagrange equation}\label{sec:EL-equation}

We start from the phase-space formulation of the problem in Eqs.~\eqref{eq:dvdPE-phase-space}--\eqref{eq:bc}. The minimisation problem of $W_{\nc}[\bm{x}]$ is constrained by the evolution equations~\eqref{eq:f1-f2-def}. This constrained minimisation problem is equivalent to minimise, without constraints, the following functional:
\begin{equation}
    J[\bm{x},\bm{p},F] = \int_0^{t_f} dt\, L^*(\bm{x},\dot{\bm{x}},\bm{p},F),
\label{eq:constrained-var-problem}
\end{equation}
where
\begin{align}
    L^*(\bm{x},\dot{\bm{x}},\bm{p},F) =& L(\bm{x}) + \bm{p}\cdot\left[\dot{\bm{x}}-\bm{f}(\bm{x};F)\right],    
\end{align}
is the new ``Lagrangian'', and $\bm{p}(t)\equiv (p_1(t),p_2(t))$ are the Lagrange multipliers, which henceforth we refer to as the momenta. We employ the usual notation for the scalar product, $\bm{u}\cdot\bm{v}=u_1 v_1+u_2 v_2$. 

The optimal path must satisfy the Euler-Lagrange equations,
\begin{subequations}
 \begin{align}
\frac{d}{dt}\left(\frac{\partial L^*}{\partial \dot{x}_i}\right) &=\frac{\partial L^*}{\partial x_i} , &  \frac{d}{dt}\left(\frac{\partial L^*}{\partial \dot{p}_i}\right) &=  \frac{\partial L^*}{\partial p_i} , \\
\frac{d}{dt}\left(\frac{\partial L^*}{\partial \dot{F}}\right) &=  \frac{\partial L^*}{\partial F}
\end{align}   
\end{subequations}
i.e. 
\begin{subequations}
 \begin{align}
      \dot{p}_1&=2\mu x_1x_2^2 + p_2(2\mu x_1x_2 +1),  \label{eq:p1dot}\\
      \dot{p}_2&=2\mu(x_1^2-1)x_2 - p_1 + \mu p_2 (x_1^2-1),  \label{eq:p2dot}\\
      \dot{x}_1 &= x_2,\label{eq:cond1} \\
      \dot{x}_2 &= -\mu(x_1^2-1)x_2 - x_1 + F, \label{eq:cond2}\\
        0 &= p_2. \label{eq:p20}
\end{align}   
\end{subequations}
Equations~\eqref{eq:cond1} and \eqref{eq:cond2} are just the evolution equations
\eqref{eq:f1-f2-def}. Equation~\eqref{eq:p20} tells us that $p_2(t)=0$, $\forall t\in [0,t_f]$, so we infer that $\dot{p}_2(t)=0$ for $t\in(0,t_f)$. Making use of Eq.~\eqref{eq:p2dot}, we get  
\begin{equation}\label{eq:p1t}
    p_1 = 2\mu(x_1^2-1)x_2, \quad \forall t\in(0,t_f).
\end{equation}
Taking time derivative of this expression and bringing to bear Eqs.~\eqref{eq:p1dot} and \eqref{eq:cond1}, we eliminate the momenta and obtain the Euler-Lagrange equation 
\begin{equation}
    (x_1^2-1)\ddot{x}_1 + x_1 \dot{x}_1^2 = 0,
\label{eq:E-L}
\end{equation}
which has to be fulfilled by the optimal path that minimises the non-conservative work. We solve this equation with the boundary conditions for $x_1$, i.e. for the position, $x_1(0)=x_{10}$, $x_1(t_f)=x_{1f}$.

At first sight, it may seem surprising that the boundary conditions for $x_2$, i.e. for the velocity $\dot{x}$, do not appear in the minimisation problem. Since $x_2=\dot{x}_1$, once we have the solution of Eq.~\eqref{eq:E-L}, we cannot tune the boundary conditions for $x_2$; the solution of the Euler-Lagrange equation will not verify the boundary condition for the velocity, in general. However, this is not problematic:  finite jumps in the velocity at the initial and final times neither change the $x$ values nor contribute to the non-conservative work. In other words, after solving the Euler-Lagrange equation \eqref{eq:E-L}, we introduce---if needed---finite jumps in the velocity at the initial and final times, of magnitude $x_{2}(0^+)-x_{20}$ and $x_{2f}-x_{2}(t_f^-)$, respectively, as explained in detail below by means of impulsive forces.

The Euler-Lagrange equations only tell us that the solution is an extremum of the considered functional, but not that it is indeed a minimum. In order to look into this issue, it is convenient to define the  ``Hamiltonian'',\footnote{Note that this definition of $H$, which is the one employed in classical mechanics, leads to the usual definition in Pontryagin's theory of optimal control.} 
\begin{equation}
    H(\bm{x},\bm{p},F)=\bm{\dot{x}}\cdot\bm{p}-L^*(\bm{x},\dot{\bm{x}},\bm{p},F)=\bm{p}\cdot\bm{f}(\bm{x};F)-L(\bm{x}).
\end{equation}
Since $H$ is linear in $F$, the following necessary condition, known as the generalised Legendre-Clebsch condition, must hold\cite{robbins_generalized_1967}
\begin{equation}
    \frac{\partial}{\partial F} 
    \left[ \frac{d^2}{dt^2} \left(\frac{\partial H(\bm{x},\bm{p};F)}{\partial F} \right)\right]\ge 0,
\end{equation}
for having a minimum of the considered functional. For our problem of concern, i.e. the minimisation of $W_{\nc}(\bm{x})$, the generalised Legendre-Clebsch condition entails that
\begin{equation}\label{eq:2nd-order-cond-min}
    x_1^2-1\ge 0. 
\end{equation}

The above condition thus restricts the solution of the minimisation problem to live in the region of phase plane for which $x_1^2\ge 1$. Therefore, we restrict ourselves to the region $|x_1|\ge 1$ in the following: in particular, both $|x_{10}|\ge 1$ and $|x_{1f}|\ge 1$, with  $\sgn(x_{10})=\sgn(x_{1f})$. For any other case, the minimisation problem would have no solution, since the condition \eqref{eq:2nd-order-cond-min} would be impossible to meet for all times. In the case $|x_{10}|\le 1$ and $|x_{1f}|\le 1$, one could, if anything, maximise the non-conservative work. This is reasonable from a physical point of view, since the non-conservative force is dissipative for $|x_1|>1$, whereas it is active (injects energy) for $|x_1|<1$.

\subsection{Optimal path and force}\label{sec:optimal-x-and-F}

Equation~\eqref{eq:E-L} has a first integral of motion, which for $x_1^2>1$ can be written as
\begin{equation}
    \dot{x}_1\sqrt{x_1^2-1} = C_1, 
    \label{eq:first-integral-EL}
\end{equation}
where $C_1$ is a  constant. We define
\begin{equation}
    g(x)\equiv \frac{1}{2}\left[x\sqrt{x^2-1}-\log\left|x+\sqrt{x^2-1}\right|\right],
\end{equation} 
such that $g'(x)=\sqrt{x^2-1}$. Therefore, $g(x_1)$ monotonically increases with $x_1$---recall that $x_1^2>1$. With this definition, the solution of Eq.~\eqref{eq:E-L} can be written as
\begin{equation}
     g(x_1(t))= C_1 t+C_0, 
    \label{eq:g-func-t}
\end{equation}
where $C_0$ is another  constant.  The constants $C_1$ and $C_0$ are calculated as functions of $x_{10}$ and $x_{1f}$,
\begin{align}\label{eq:C0-C1}
    C_0 &= g(x_{10}), & C_1=\frac{g(x_{1f})-g(x_{10})}{t_f},
\end{align}
which completes the solution for the optimal trajectory that minimises the non-conservative work.
The initial and final velocities over the optimal trajectory $\dot{x}_1(0^+)$ and $\dot{x}_1(t_f^-)$ are obtained from Eq.~\eqref{eq:first-integral-EL}:
\begin{align}
   x_2(0^+)=\dot{x}_1(0^+)=&\frac{C_1}{\sqrt{1-x_{10}^2}}, & x_2(t_f^-)=\dot{x}_1(t_f^-)=&\frac{C_1}{\sqrt{1-x_{1f}^2}}. 
\end{align}
Note that, as already anticipated after Eq.~\eqref{eq:E-L}, the velocity $x_2$ does not comply in general with the boundary conditions but this is not problematic. We can introduce two impulsive contributions to the force---i.e. two delta peaks---at the initial and final times to fix this issue, neither changing the particle position nor performing any work.

Making use of the driven dvdPE~\eqref{eq:driven-vdP}, particularised for the optimal trajectory, we get the driving force 
\begin{equation}\label{F-opt-func-t-EL}
    F_{\EL}(t) = -C_1^2 \,\frac{ x_1(t)}{[x_1^2(t)-1]^2}+\mu\, C_1 \sqrt{x_1^2(t)-1}+x_1(t),
\end{equation}
for $0^+<t<t_f^-$. At $t=0^+$ and $t=t_f^-$,  the finite jumps in $x_2=\dot{x}_1$ entail that $F(t)$ has delta peaks, at $t=0^+$ and $t=t_f^-$, a behaviour that has also been found in the optimisation of the irreversible work done on a harmonically confined Brownian particle in the underdamped case.\cite{gomez-marin_optimal_2008} More specifically, we have for the optimal driving
\begin{align}\label{F-opt-func-t-EL-with-deltas}
    F_{\opt}(t)= F_{\EL}(t)&+\left[x_2(0^+)-x_{20}\right] \delta(t-0^+) \nonumber \\
    &
    +\left[x_{2f}-x_2(t_f^-)\right] \delta(t-t_f^-).  
\end{align}


\subsection{Transversality condition for variable endpoint}\label{sec:transvers-cond-var-endpoint}

Now we take into account that the final point is not fixed, we only know that it belongs to the limit cycle. Therefore, in the variational procedure, $\delta x_{1f}$ and $\delta x_{2f}$ do not vanish and the following condition must hold:
\begin{equation}
    0 = \left.\left(\frac{\partial L^*}{\partial \dot{x}_1}\delta x_{1} + \frac{\partial L^*}{\partial \dot{x}_2}\delta x_{2}\right)\right|_{t=t_f} = \bm{p}_f \cdot \delta\bm{x}_f,
\label{eq:transvers-cond}
\end{equation}
where $p_{1f}\equiv p_1(t_f)$, $p_{2f}\equiv p_2(t_f)$.
Equation~\eqref{eq:transvers-cond} is known as the transversality condition, since it tells us that the vectors $\bm{p}_f$ and $\delta\bm{x}_f$ are orthogonal{---see also Appendix~\ref{app-transvers-cond}.} 

Since the final point belongs to the limit cycle, Eq.~\eqref{eq:fLC-phase-space} implies that $\delta \bm{x}_f$ is parallel to the tangent vector to the phase plane trajectory for the undriven van der Pol equation,
\begin{equation}\label{eq:deltaxf-tangent}
    \delta\bm{x}_f \parallel (f_1(x_{2f}),f_2(x_{1f},x_{2f};F=0)).
\end{equation}
Recalling that $p_2(t)=0$, $\forall t$, the transversality condition tells us that $p_{1f}f_1(x_{2f})=p_{1f} x_{2f}=0$. Since $p_1(t)$ is a continuous function of time, $p_{1f}=p_1(t_f^-)$, and making use of Eq.~\eqref{eq:p1t} we have
\begin{equation}\label{eq:p1tf-trans}
    2\mu(x_1^2(t_f^-)-1)x_2(t_f^-) x_{2f}=0.
\end{equation}
Note that, as discussed before, the velocity is in general discontinuous at the final time, in general $x_2(t_f^-)\ne x_{2f}$.

There appear several possibilities:
\begin{enumerate}[label=T\arabic*]
    \item\label{item1}\!\!. $x_{2f}=0$: This condition is only fulfilled at the leftmost and rightmost points of the limit cycle, i.e. $x_{1f}=\pm x_{\LC}^{\max}$. 
    {
    \item\label{item2}\!\!. $x_2(t_f^-)=0$: 
    This condition entails, together with Eq.~\eqref{eq:cond1} and Eq.~\eqref{eq:first-integral-EL}, that $C_1=0$. This implies that $x_2(t) = 0$ for $t\in(0,t_f)$  In other words, $x_1(t)$ is constant, $x_{10}=x_{1f}$. Clearly, this possibility only makes sense if $x_{10}\in[-x_{\LC}^{\max},-1]\cup[+1,+x_{\LC}^{\max}]$.}
    \item\label{item3}\!\!. $x_{1f}=x_1(t_f^-)=1$: Again, Eq.~\eqref{eq:first-integral-EL} implies that $x_1(t)$ is constant, $x_{10}=x_{1f}=1$, so this possibility is included in \ref{item2}. 
\end{enumerate}

\subsection{Minimum non-conservative work}\label{sec:min-non-conserv-work}

Taking now into account Eqs.~\eqref{E-and-Wnc-defs}, \eqref{eq:first-integral-EL}, and \eqref{eq:C0-C1}, the non-conservative work $W_{\nc}^{\min}$ along the Euler-Lagrange path for a given final point $x_{1f}$ is 
\begin{align}\label{eq:Wnc-min}
    W_{\nc}^{\min} &= \mu \, C_1^2 \, t_f = \mu \frac{\left[g(x_{1f})-g(x_{10})\right]^2}{4t_f}.    
\end{align}
To obtain the minimum value of the work, we employ the result obtained from  the transversality condition: there are two candidates for the optimal endpoint, (i) $x_{1f}=\pm x_{\LC}^{\max}$ and (ii) $x_{1f}=x_{10}$. First, if the point $(x_{10},0)$ lies inside the limit cycle, i.e. $|x_{10}|\le x_{\LC}^{\max}$,  we have that the minimum value of $W_{\nc}$ is reached for $x_{1f}=x_{10}$, for which $W_{\nc}^{\min}$ vanishes. Second, if the point $(x_{10},0)$ lies outside  the limit cycle, i.e. $|x_{10}|>x_{\LC}^{\max}$, we have that the minimum value of $W_{\nc}$ is reached at $x_{1f}=\sgn(x_{10}) x_{\LC}^{\max}$. 

Therefore, the optimal final position over the limit cycle---in terms of non-conservative work---is the one ``closest in $x$'' to the initial position $x_{10}$:
\begin{equation}\label{eq:xf-opt}
 x_{1f}^{\opt} =
\begin{cases}
    x_{10},  &\text{if } b \leq |x_{10}|\leq x_{\LC}^{\max},\\
     \sgn (x_{10})x_{\LC}^{\max},  &\text{if } |x_{10}|>x_{\LC}^{\max}.
\end{cases}
\end{equation}
{ We have defined $b=1$ as the position at which the damping force vanishes.} We recall that we have assumed  $|x_{10}|\ge 1$, $|x_{1f}|\ge 1$ to minimise $W_{\nc}$. Finally, the minimum non-conservative work to an arbitrary point of the limit cycle is
\begin{equation}\label{eq:W-min-unbounded}
 W_{\nc}^{\min} =
\begin{cases}
    0,  &\text{if } b\leq|x_{10}|\leq x_{\LC}^{\max},\\
       \mu \dfrac{\left[g(x_{\LC}^{\max})-g(|x_{10}|)\right]^2}{4t_f},  &\text{if } |x_{10}|>x_{\LC}^{\max}.
\end{cases}
\end{equation}
The final expression for the minimum non-conservative work \eqref{eq:W-min-unbounded} suggests the introduction of a scaled connection time
\begin{equation}
    \sctf \equiv t_f/\mu,
\end{equation}
so that $W_{\nc}^{\min}$ can be written as
\begin{equation}\label{eq:W-min-unbounded-v2}
    W_{\nc}^{\min}=  \frac{\left[g(x_{\LC}^{\max})-g(|x_{10}|)\right]^2}{4 \sctf} H(|x_{10}|-x_{\LC}^{\max}), \quad \forall\mu,
\end{equation}
where $H(x)$ is Heaviside's step function, $H(x)=1$, $\forall x>0$, $H(x)=0$, $\forall x\le 0$. In terms of the scaled time $\sctf$, $W_{\nc}^{\min}$ still depends on the damping coefficient $\mu$ through $x_{\LC}^{\max}$. This ``residual'' dependence on $\mu$ disappears in the small damping limit; making use of Eq.~\eqref{eq:xlc-max-0}, we get
\begin{equation}\label{eq:W-min-unbounded-small-mu}
    \widetilde{W}_{\nc}^{\;\min} = \frac{\left[g(\widetilde{x}_{\LC}^{\;\max})-g(|x_{10}|)\right]^2}{4 \sctf} H(|x_{10}|-\widetilde{x}_{\LC}^{\;\max}), \quad \mu\ll 1.
\end{equation}
{ We recall that $x_{\LC}^{\max}$ is very close to $\widetilde{x}_{\LC}^{\;\max}=2$ for all $\mu$, the deviation from it being always below $3\%$.\cite{turner_maximum_2015} Therefore, in the following subsection of physical interpretation of the results, we mainly focus on Eq.~\eqref{eq:W-min-unbounded-small-mu}. }

\subsection{Physical interpretation of the results}\label{sec:phys-interpret-1}

\begin{figure}
    \centering
    \includegraphics[width=3in]{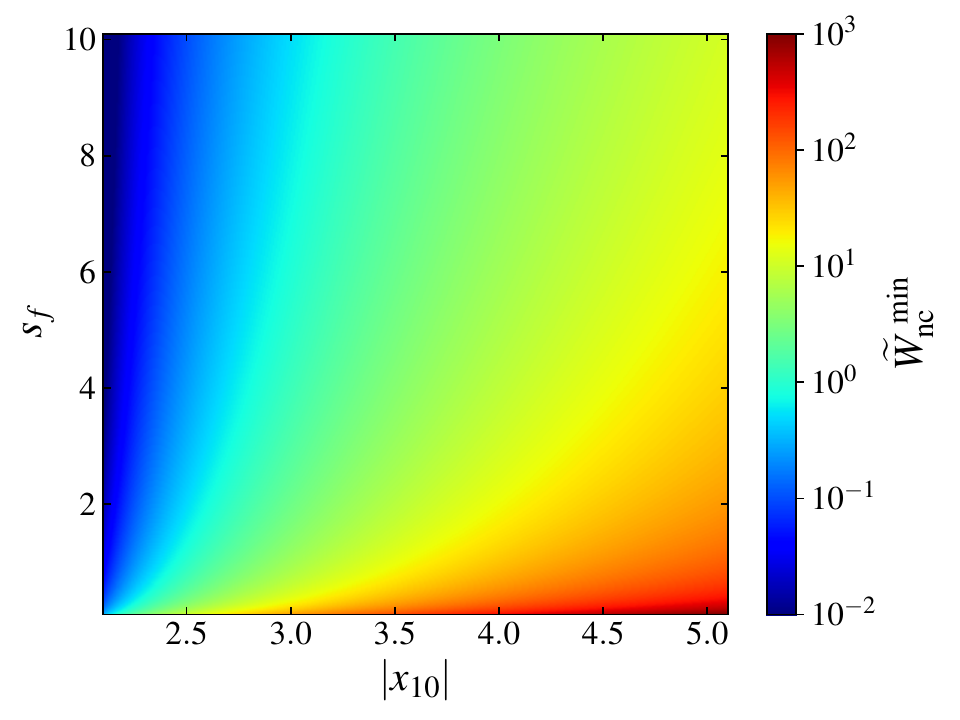} 
    \caption{\label{fig:Wnc-dens-plot} Minimum non-conservative work $\widetilde{W}_{\nc}^{\;\min} $ as a function of the scaled connection time $s_f$ and the initial position $|x_{10}|$. The plotted data correspond to the theoretical expression~\eqref{eq:W-min-unbounded-small-mu} { for the small damping limit, which is independent of $\mu$.} It is clearly observed how $\widetilde{W}_{\nc}^{\;\min}$ decreases as $s_f$ increases and $|x_{10}|$ decreases. }
\end{figure}

The minimum non-conservative work $\widetilde{W}_{\nc}^{\;\min}$ depends on the initial point $|x_{10}|$ and the scaled connection time $s_f=t_f/\mu$. This is illustrated with the density plot of {$\widetilde{W}_{\nc}^{\;\min}$} in Fig.~\ref{fig:Wnc-dens-plot}. We only plot the region $|x_{10}|\ge \widetilde{x}_{\LC}^{\;\max}$, because $\widetilde{W}_{\nc}^{\;\min}$ identically vanishes for $|x_{10}|<\widetilde{x}_{\LC}^{\;\max}$.\footnote{Due to the symmetry $x_1\to -x_1$ of our problem, all the plots in this section are for $|x_{10}|\ge 1$.} A first physical consequence of our result~\eqref{eq:W-min-unbounded-small-mu} is that $\widetilde{W}_{\nc}^{\;\min}$ is proportional to the inverse of the connection time, $\widetilde{W}_{\nc}^{\;\min}\propto s_f^{-1}$. In the limit of small damping { we are focusing on}, the proportionality constant depends only on $|x_{10}|$.\footnote{For an arbitrary value of $\mu$, the proportionality constant also depends { very weakly} on the damping coefficient $\mu$ through $x_{\LC}^{\max}$.} 
This scaling with the connection time is similar to the one found for the minimum irreversible work for the accelerated connection between equilibrium states of mesoscopic system in stochastic thermodynamics.\cite{aurell_optimal_2011,sivak_thermodynamic_2012,plata_optimal_2019,zhang_work_2020,plata_taming_2021,guery-odelin_driving_2023} 

\begin{figure}
    \centering
    \includegraphics[width=3.25in]{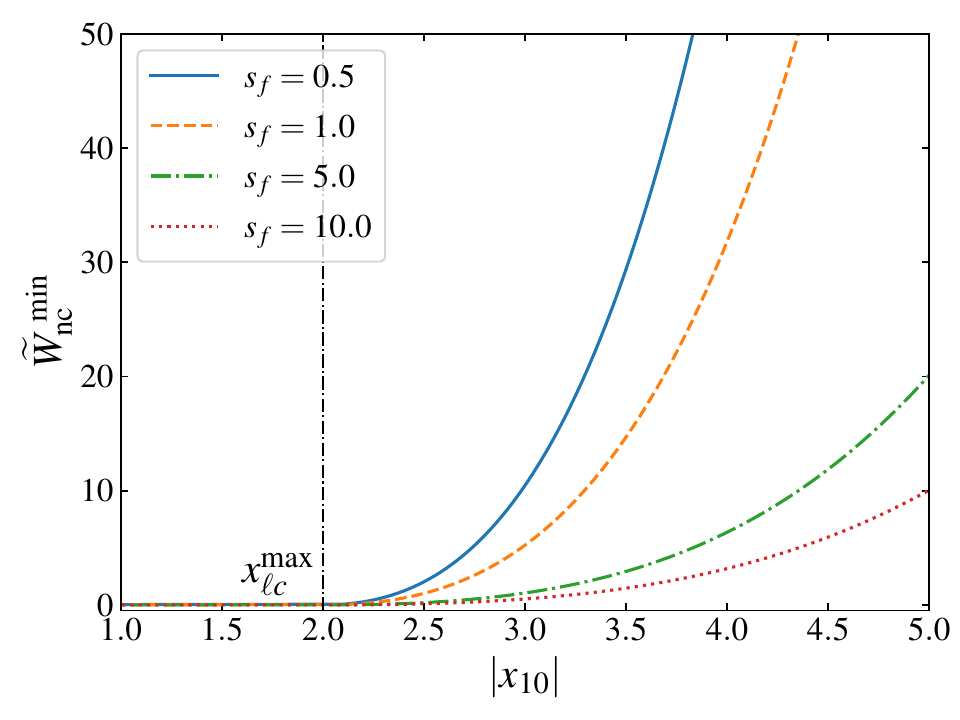}
    \caption{\label{fig:WncVSsfx0} Non-conservative minimum work $\widetilde{W}_{\nc}^{\;\min} $ as a function of  $|x_{10}|$. Similarly to Fig.~\ref{fig:Wnc-dens-plot}, the plotted data correspond to the theoretical expression~\eqref{eq:W-min-unbounded-small-mu}. It can be observed that $\widetilde{W}_{\nc}^{\;\min} $ increases from zero, for $|x_{10}|\le \widetilde{x}_{\LC}^{\;\max}$, to infinity, in the limit $|x_{10}|\to +\infty$.}
 \end{figure}

The dependence of the rhs of {$\widetilde{W}_{\nc}^{\;\min}$} on the initial position $x_{10}$ is illustrated in 
Fig.~\ref{fig:WncVSsfx0}, for several values of the scaled connection time $s_f$. For $|x_{10}|\leq \widetilde{x}_{\LC}^{\;\max}$, i.e. when the initial position lies between the extremal positions $\pm \widetilde{x}_{\LC}^{\;\max}$ of the limit cycle, $W_{\nc}^{\min}$ identically vanishes.  For $|x_{10}|>\widetilde{x}_{\LC}^{\;\max}$, i.e. when the initial position lies outside the extremal positions of the limit cycle, $W_{\nc}^{\min}\ne 0$ and increases with $|x_{10}|$, specifically as $[g(x_{10})-g(\widetilde{x}_{\LC}^{\;\max})]^2$. 

Let us analyse now the optimal trajectory and the optimal driving force that lead to the minimum non-conservative work. On the one hand, for $|x_{10}|\le x_{\LC}^{\max}$, $x_{1f}^\opt =x_{10}$ and the optimal trajectory leading to the limit cycle corresponds to $C_1=0$. Therefore, $x_2=\dot{x}_1$ identically vanishes over the optimal solution for $0^+<t<t_f^-$, the system remains at rest at the point $(x_{10},0)$ of phase plane. The optimal driving force, as given by Eq.~\eqref{F-opt-func-t-EL}, is constant, $F_\EL(t)=x_{10}$. On the other hand, for $|x_{10}|> x_{\LC}^{\max}$, $x_{1f}^{\opt}=x_{\LC}^{\max}$ and thus $C_1\ne 0$. Since the sign of $x_2$, i.e. the sign of $\dot{x}_1$, does not change with time and is always that of $C_1$, as given by Eq.~\eqref{eq:first-integral-EL}, $x_1(t)$ monotonically varies between $x_{10}$ and $x_{1f}$. In this case, the optimal driving force ~\eqref{F-opt-func-t-EL} is not constant. In both cases, the initial and final values of the velocity are adjusted as we have already commented below Eq.~\eqref{F-opt-func-t-EL}: at $t=0^+$ and $t=t_f^-$ we introduce two finite-jump discontinuities in the velocity. We recall that this finite jump discontinuities in the velocity produce no non-conservative work, although they certainly entail that the driving force has delta-peaks at the initial and final times, as given by Eq.~\eqref{F-opt-func-t-EL-with-deltas}. { Note that the only role of the initial velocity $x_{20}$ in the optimal protocol is to modify the correct amplitude for the impulsive contributions to the force, since the Euler-Lagrange solution for $0<t<t_f$ is blind to the boundary values of $x_2$.}

\begin{figure}
    \centering
    \includegraphics[width=3in]{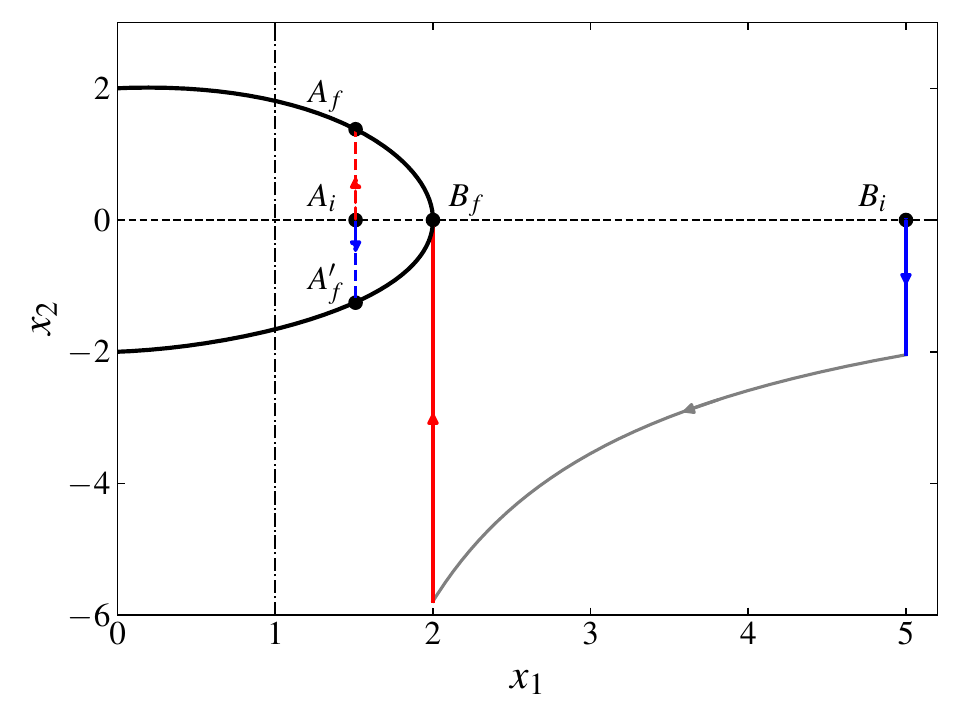}
    \includegraphics[width=3.07in]{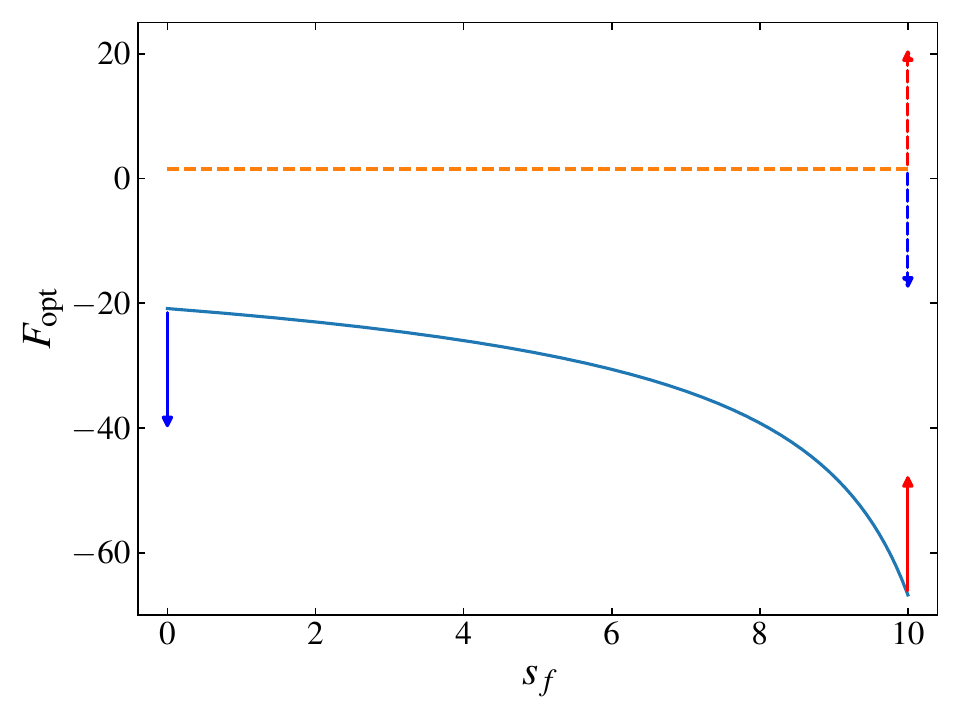}
    \caption{\label{fig:ELpathFunbounded} Optimal phase plane trajectory (top) and force (bottom) for minimising the non-conservative work. Specifically, a damping constant $\mu=0.1$ and a scaled connection time $s_f=10$ have been used. {This corresponds to an acceleration factor $t_R/t_f=40$.} Two different values of the initial phase plane point, namely $A_i=(1.5,0)$ (dashed lines) and $B_i=(5,0)$, are considered. { The corresponding final points are $A_f\simeq(1.5,1.4)$, $A'_f\simeq(1.5,-1.3)$ and $B_f\simeq(2,0)$, respectively.} In the top panel, the dot-dashed vertical line marks the limit of the phase plane region in which the non-conservative work is minimised, i.e. $|x_{10}|>1$.  
    }
\end{figure}

Figure~\ref{fig:ELpathFunbounded} shows the optimal trajectory on the phase plane (top panel) and the corresponding optimal force (bottom panel). Specifically, we have employed a damping coefficient $\mu=0.1$ and a scaled connection time $s_f=10$, and two initial points on the phase plane: $A_i=(1.5,0)$ and $B_i=(5,0)$, which correspond to the two cases discussed in the previous paragraph. For point $A_i$ (dashed lines), there are two equivalent possibilities {for the final point, $A_f$ and $A'_f$,} which correspond to driving the system towards the point of the limit cycle just above (red line) or below (blue line) $A_i$. For point $B_i$, only one optimal trajectory is possible, that ending on the phase plane point $B_f=(x_{\LC}^{\max},0)$.\footnote{For $\mu=0.1$, $x_{\LC}^{\max}=2.00010$.\cite{turner_maximum_2015}} In the top panel, the arrows mark the direction of the movement on the phase plane---recall that $x_2=\dot{x}_1$. In the bottom panel, the delta peaks of the force at the initial and final times are marked with the vertical arrows, which indicate the sign of the impulsive forces. 

Equation \eqref{eq:W-min-unbounded-v2} entails the emergence of a trade-off relation between connection time $t_f$ and non-conservative work for an accelerated synchronisation to the limit cycle. Taking into account that $W_{\nc}[\bm{x}]\geq W_{\nc}^{\min}$, we have in general that
\begin{equation}
    \sctf W_{\nc}[\bm{x}]\geq\frac{\left[g(x_{\LC}^{\max})-g(|x_{10}|)\right]^2}{4} H(|x_{10}|-x_{\LC}^{\max}), \quad \forall\mu.
\end{equation}
The monotonic decreasing behaviour of $W_{\nc}^{\min}$ as a function of the connection time entails that  there appears a minimum time for the connection, for a given value of the non-conservative work $W_{\nc}$. That is, a speed limit arises:
\begin{equation}
    \sctf \ge \sctf^{\min}=\frac{\left[g(x_{\LC}^{\max})-g(|x_{10}|)\right]^2}{4W_{\nc}} H(|x_{10}|-x_{\LC}^{\max}).
\end{equation}
In the small damping limit, we have
\begin{equation}
    \sctf W_{\nc}[\bm{x}]\geq \frac{\left[g(\widetilde{x}_{\LC}^{\;\max})-g(|x_{10}|)\right]^2}{4} H(|x_{10}|-\widetilde{x}_{\LC}^{\;\max}), \; \mu\ll 1.
\end{equation}
The bound on the rhs is independent of $\mu$, it only depends on the initial condition $x_{10}$ and remains of the order of unity as long as $|x_{10}|-\widetilde{x}_{\LC}^{\;\max}=O(1)$.


\section{Generalisation of the results}\label{sec:general-results}

\subsection{Liénard equation}\label{sec:lienard-eq}

The van der Pol equation is a particular case of the Liénard equation,\cite{iacono_class_2011,messias_time-periodic_2011,ghosh_lienard-type_2014,shah_conservative_2015,turner_maximum_2015,gine_lienard_2017,mishra_chimeras_2023,strogatz_nonlinear_2024} which we write as
\begin{equation}\label{eq:lienard}
    \ddot{x}+\mu h(x)\dot{x}+V'(x)=0,
\end{equation}
where $h(x)$ and $V(x)$ are even functions of $x$, $h(x)\in\mathcal{C}^1$ and $V(x)\in\mathcal{C}^2$. This equation describes a particle moving under the action of a conservative force $-V'(x)$, stemming from a potential energy $V(x)$, and a non-conservative non-linear viscous force $-\mu h(x)\dot{x}$. Equation~\eqref{eq:lienard} has a unique stable limit cycle under the following assumptions: (i) $V(x)$ is a confining potential with only one minimum at $x=0$, (ii) the function 
\begin{equation}
    \xi(x)=\int_0^x dx' h(x')
\end{equation}
has the properties: (a) $\xi(x)$ has only one positive zero at $x=a$, with $\xi(x)<0$ for $0<x<a$ and $\xi(x)>0$ for $x>a$, and (b) $\xi(x)$ is non-decreasing for $x>a$, with $\lim_{x\to +\infty}\xi(x)=+\infty$.\footnote{For the van der Pol equation, $\xi(x)=(x^3-3x)/3$ and $a=\sqrt{3}$.}

Remarkably, most of the results for the minimisation of the non-conservative work, as derived in Sec.~\ref{sec:minimisation-nc-work-unbounded}, also apply to the more general Liénard equation, with small changes. For the Liénard equation, the region in which the non-conservative work can be minimised is defined by the condition $h(x)\ge 0$. Therein, the minimum non-conservative work for a fixed value of $x_{f}$ is still given by Eq.~\eqref{eq:Wnc-min}, defining $g(x)$ as the primitive of $\sqrt{h(x)}$, i.e. $g'(x)=\sqrt{h(x)}$.  Also, the optimal final point is given by Eq.~\eqref{eq:xf-opt}, with $x_{\LC}^{\max}$ being the maximum value of $x$ over the limit cycle. Moreover, this entails that the emergence of a speed limit extends to the Liénard equation. Below we summarise how these results are derived.

Our starting point is the driven Liénard equation, i.e. we add a force $F(t)$ that we control to its rhs:
\begin{equation}\label{eq:driven-lienard}
    \ddot{x}+\mu h(x) \dot{x}+V'(x)=F(t),
\end{equation}
or
\begin{equation}\label{eq:driven-lienard-phase-space}
    \dot{x}_1=\underbrace{x_2}_{f_1}, \quad \dot{x}_2=\underbrace{-\mu h(x_1)x_2-V'(x_1)+F}_{f_2}.
\end{equation}
We assume that $V(x)$ and $h(x)$ verify the conditions under which the undriven Liénard equation has a unique stable limit cycle.  Again, we consider the optimal synchronisation to this limit cycle, in terms of the minimum non-conservative work. The work done by the non-conservative viscous force is 
\begin{equation}\label{eq:Wnc-lienard}
    W_{\nc}[\bm{x}]=\mu \int_0^{t_f} dt\, h(x_1)\, x_2^2 .
\end{equation}
Once more, we would like to minimise the non-conservative work for the synchronisation to the limit cycle, i.e. we would like to minimise the functional in Eq.~\eqref{eq:Wnc-lienard} with the boundary conditions \eqref{eq:bc}---with $\chi_{\LC}(\bm{x},\mu)=0$ standing now for the limit cycle of the Liénard equation.

The above minimisation problem can be tackled in the same way as for the van der Pol case, as a variational problem with constraints that can be introduced with the introduction of Lagrange multipliers, the momenta $\bm{p}$. Due to the linearity of $f_2$ in the force, 
\begin{equation}\label{eq:p1t-p2t-lienard}
   p_2(t)=0, \qquad  p_1(t)=2\mu h(x_1)x_2, \quad \forall t\in(0,t_f).
\end{equation}
Taking into account this, we get the Euler-Lagrange equation for the optimal trajectory that minimises $W_{\nc}$,
\begin{equation}\label{eq:EL_lienard}
    2 h(x_1) \ddot{x}_1+h'(x_1) \dot{x}_1^2=0.
\end{equation}
The necessary Clebsch-Legendre condition for having a minimum is now
\begin{equation}\label{eq:hx1>0}
    h(x_1)\ge 0.
\end{equation}
Therefore, for the Liénard equation, we restrict ourselves to the region of phase plane in which $h(x_1)>0$. For the sake of simplicity, we consider that there is only one point, $x_1=b$, at which $h(x_1)=0$ for $x_1>0$. Therefore, Eq.~\eqref{eq:hx1>0} is equivalent to $|x_1|\ge b$; for the van der Pol equation, $b=1$.

Again, we have a  constant of motion,
\begin{equation}\label{eq:first-integ-lienard}
\dot{x}_1\sqrt{h(x_1)}=C_1.
\end{equation}
Particularising this equation to the van der Pol oscillator, where $h(x)=x^2-1$, reproduces Eq.~\eqref{eq:first-integral-EL} of Sec.~\ref{sec:minimisation-nc-work-unbounded}. In an analogous way, we find the solution for the optimal trajectory:
\begin{equation}
    g'(x)= \sqrt{h(x)} \implies g(x_1)=C_1 t+C_0.
\end{equation}
Therefore, Eqs.~\eqref{eq:g-func-t} and \eqref{eq:C0-C1} of Sec.~\ref{sec:minimisation-nc-work-unbounded} hold with our redefinition of $g(x)$. The corresponding driving force for the optimal path is
\begin{equation}
    F_{\EL}(t)=-C_1^2 \frac{h'(x_1)}{2 h^2(x_1)}-\mu C_1 \sqrt{h(x_1)}+x_1,
\end{equation}
for $0^+ < t< t_f^-$. At the initial and final times, two delta peaks are necessary to reach the target values for  the initial and final velocities, $x_{20}$ and $x_{2f}$, as described by Eq.~\eqref{F-opt-func-t-EL-with-deltas}.

The transversality condition selects the final point over the limit cycle that gives the minimum non-conservative work. The whole discussion in Sec.~\ref{sec:transvers-cond-var-endpoint} applies to the Liénard equation, since the explicit shape of the limit cycle was not employed. The unique equation that is specific to the van der Pol equation is \eqref{eq:p1tf-trans}, since we made use of the particular expression for $p_1(t)$. Yet, the conclusion obtained from it is valid, since for the Liénard equation 
\begin{equation}\label{eq:p1tf-Lienard}
  p_1(t_f^-)=2\mu h(x_1(t_f^-))x_2(t_f^-),  
\end{equation}
and therefore the transversality condition entails that either $x_2(t_f^-)$ or $x_{2f}$ vanishes.\footnote{Similarly to the discussion below Eq.~\eqref{eq:p1tf-trans}, the case $h(x_1(t_f^-))=0$ is a particular case of $x_2(t_f^-)=0$.}

Equation~\eqref{eq:Wnc-min} for the minimum non-conservative work for a fixed final point $x_{1f}$ immediately holds, since Eq.~\eqref{eq:first-integ-lienard} entails that the integrand of $W_{\nc}$ equals $\mu C_1^2$ over the optimal trajectory. Our discussion on the transversality condition above entails that $W_{\nc}$ takes its minimum value at the point $x_{1f}^{\opt}$ given by Eq.~\eqref{eq:xf-opt}, and  Eqs.~\eqref{eq:W-min-unbounded}--\eqref{eq:W-min-unbounded-v2} also hold for the driven Liénard equation.

The above analysis implies that the scaled time $s=t/\mu$ is also the relevant timescale for the optimal synchronisation to the limit cycle in the Liénard equation. In the small damping limit $\mu\ll 1$, the approximate value of the rightmost point of the limit cycle $\widetilde{x}_{\LC}^{\;\max}$ can also be obtained by energetic arguments, similarly to those employed for the van der Pol equation: the limit cycle corresponds to a closed orbit $\frac{1}{2}x_2^2+V(x_1)=E^{(0)}$, where $E^{(0)}$ is determined by imposing that the non-conservative work vanishes over the closed orbit.


\subsection{Minimisation of the total work}\label{sec:min-total-work}

Let us consider the minimisation of the total work $W$ done by the external force $F(t)$, not only the non-conservative contribution $W_{\nc}$. More precisely, we consider a fixed initial point $(x_{10},x_{20})$ on the phase plane and look for (i) the phase plane trajectory $(x_1(t),x_2(t))$ verifying the evolution equations \eqref{eq:driven-lienard-phase-space} and (ii) the final point  $(x_{1f},x_{2f})$ over the limit cycle that minimises the total work. Similarly to Eq.~\eqref{eq:W-def}, we have that
\begin{equation}\label{eq:total-work}
    W=E(\bm{x}_f)-E(\bm{x}_0)+W_{\nc}[\bm{x}], \text{ with } E(\bm{x})\equiv\frac{1}{2}x_2^2+V(x_1).
\end{equation}

We consider an infinitesimal variation of the phase plane path $\delta\bm{x}$ around the optimal one and impose that the variation of $E(\bm{x}_f)+W_{\nc}[\bm{x}]$, with the constraints given by the evolution equations \eqref{eq:driven-lienard-phase-space} incorporated with the Lagrange multipliers $\bm{p}(t)$. The ``bulk'' term of the variation, for  $t\in (0,t_f)$, leads to the Euler-Lagrange equations derived in Sec.~\ref{sec:EL-equation}. For the optimal path verifying the Euler-Lagrange equation \eqref{eq:EL_lienard}, only the variation at the upper limit survives and must vanish:
\begin{equation}\label{eq:extended-transvers}
    \left[\nabla_{\bm{x}_f} E(\bm{x}_f)+\bm{p}_f\right] \cdot \delta\bm{x}_f=0,
\end{equation}
where $\nabla_{\bm{x}} E(\bm{x})=(V'(x_1),x_2)$, { see Appendix~\ref{app-transvers-cond} for details.} Therefore, the \textit{modified transversality condition} \eqref{eq:extended-transvers} tells us that the vector $\nabla_{\bm{x}_f} E(\bm{x}_f)+\bm{p}_f$ is perpendicular to $\delta\bm{x}_f$, which verifies the paralellism condition in Eq.~\eqref{eq:deltaxf-tangent}. Then we have
\begin{align}\label{eq:extended-transvers-2}
    0&=\left[\cancel{V'(x_{1f})}+p_{1f}\right]x_{2f}+x_{2f}\left[-\mu h(x_1)x_{2f}-\cancel{V'(x_{1f})}\right] \nonumber \\
    &=x_{2f}\left[p_1(t_f^-)-\mu h(x_1)x_{2f}\right] 
    = \mu h(x_{1f})x_{2f}\left[2 x_2(t_f^-)-x_{2f}\right].
\end{align}

Equation~\eqref{eq:extended-transvers-2} has relevant implications. There appear three possibilities:
\begin{enumerate}[label=MT\arabic*]
    \item\!\!. $x_{2f}=0$: This condition is only fulfilled at the leftmost and rightmost points of the limit cycle, i.e. $x_{1f}=\pm x_{\LC}^{\max}$.\label{item1-lienard} 
    \item\!\!. $x_2(t_f^-)=x_{2f}/2$: 
    Making use of the first integral~\eqref{eq:first-integ-lienard}, together with Eq.~\eqref{eq:C0-C1}, we get
    \begin{equation}\label{eq:possibility2-total-work}
        x_{2f}=\frac{2[g(x_{1f})-g(x_{10})]}{t_f \sqrt{h(x_{1f})}},
    \end{equation}
    provided that $h(x_{1f})\ne 0$ (see \ref{item3-lienard}).\label{item2-lienard}
    \item\!\!. $h(x_{1f})=0$, i.e. $x_{1f}=b$: The first integral~\eqref{eq:first-integ-lienard} tells us that $x_1(t)$ is constant, $x_1(t)=b$, $\forall t\in[0,t_f]$.\label{item3-lienard}
\end{enumerate}
\ref{item1-lienard}--\ref{item3-lienard} are the conditions for the minimisation of the total work $W$ corresponding to \ref{item1}--\ref{item3} for the minimisation of the non-conservative work. Note that \ref{item3-lienard} is no longer a particular case of \ref{item2-lienard}. 

It is interesting to analyse the modified transversality conditions \ref{item1-lienard}--\ref{item3-lienard} in the limit of small damping $\mu\ll 1$, in which the relevant timescale for the connection is $s_f=t_f/\mu$. Therein, Eq.~\eqref{eq:possibility2-total-work} for \ref{item2-lienard} is rewritten as
\begin{equation}
    \mu\, x_{2f}=\frac{2[g(x_{1f})-g(x_{10})]}{s_f \sqrt{h(x_{1f})}}.
    \label{eq:WtotalTransVer2}
\end{equation}
Note that this is a closed equation for $x_{1f}$ after considering that the final point $(x_{1f},x_{2f})$ belongs to the limit cycle, i.e. $\chi_{\LC}(x_{1f},x_{2f})=0$. Below we study the cases corresponding to positions inside the limit cycle, $b\le |x_{10}|\le x_{\LC}^{\max}$, and outside the limit cycle, $|x_{10}|> x_{\LC}^{\max}$, separately.

First, we consider that the initial position lies inside the limit cycle, $b\le |x_{10}|\le x_{\LC}^{\max}$. Assuming that $x_{2f}=O(1)$,  we have that $x_{1f}-x_{10}=O(\mu)$ for $s_f=O(1)$, i.e. we recover to the lowest order the solution for the non-conservative work. This is reasonable from a physical point of view: for $\mu\ll 1$, the energy is approximately constant---with $O(\mu)$ corrections---over the limit cycle, so minimising the non-conservative work and the total work is equivalent. (The non-conservative work is of the order of unity for any non-vertical trajectory.) This continues to be true for very short connection times $s_f\ll 1$. Again, this is reasonable: for very short connection times, the non-conservative work diverges for any non-vertical trajectory and dominates the minimisation of the total work. The only exception is thus the regime of long connection times $s_f\gg 1$, which includes order of unity times in the original timescale: note that $t_f=1$ gives $\mu s_f=O(1)$, which would lead to $x_{1f}-x_{10}=O(1)$. For $\mu s_f \gg 1$, $x_{1f}\to b$---recall that $h(x_1)$ vanishes at $x_1=b$.

\begin{figure}
    \centering
    \includegraphics[width=3in]{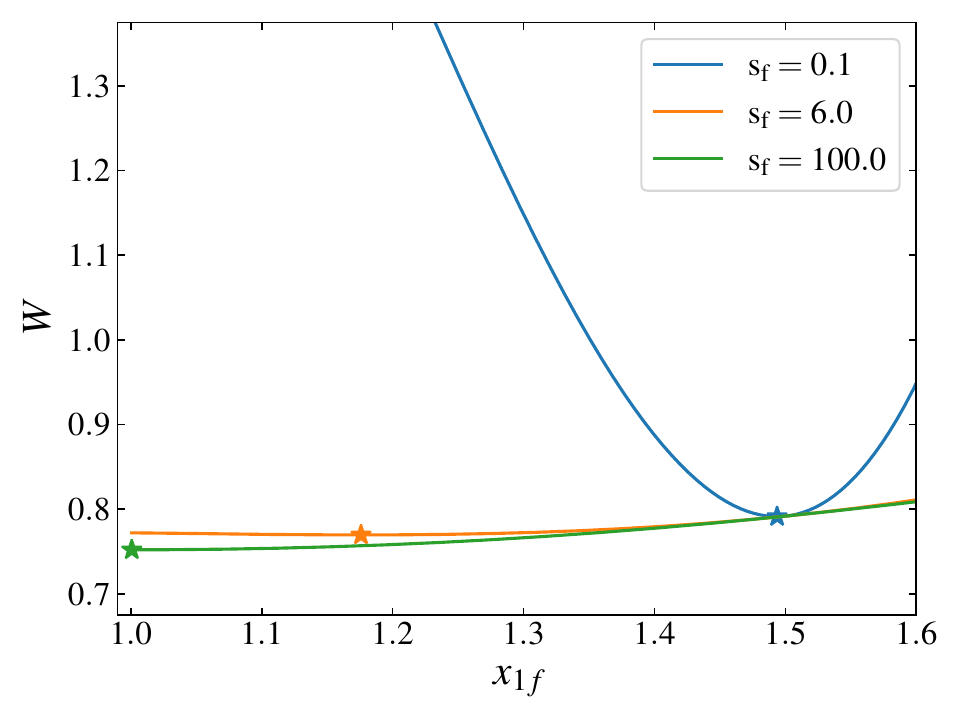}
    \hspace*{-0.2cm}\includegraphics[width=3in]{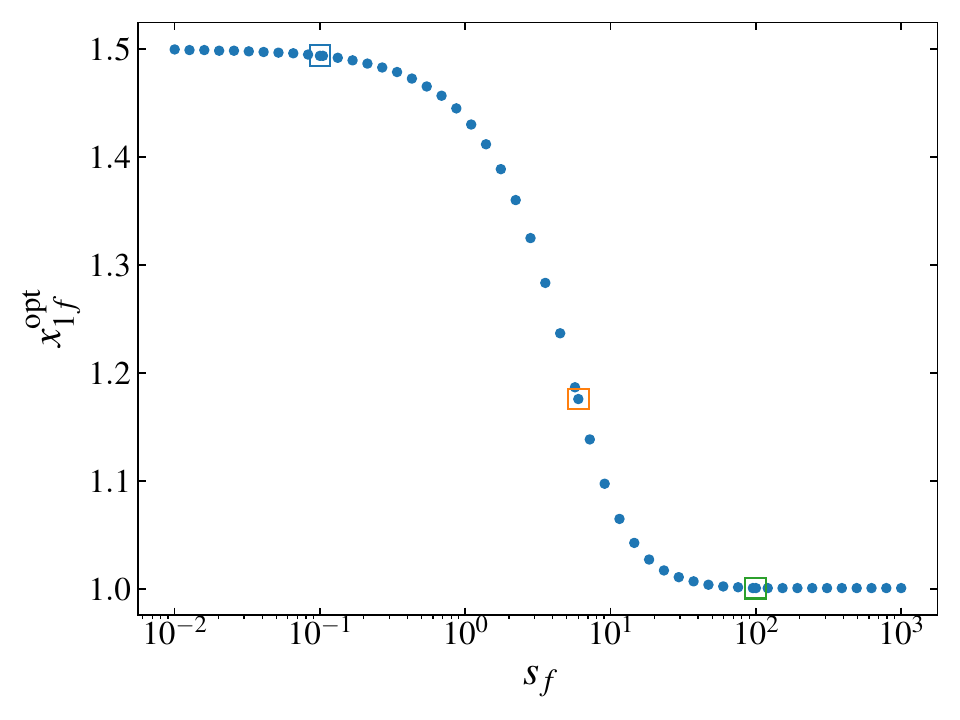}
    \caption{{Total work $W$ as a function of the final point over the limit cycle $x_{1f}$ (top) and optimal final point $x_{1f}^{\opt}$ as a function of the scaled time $s_f$ (bottom). In particular, the plotted curves correspond to the van der Pol oscillator with a damping constant $\mu=0.1$, and an initial point inside the limit cycle, namely $x_{10}=1.5$. On the top panel, the stars mark the points at which $W$ reaches its minimum as a function of $x_{1f}$, for the considered values of $s_f$ shown in the legend. These points are singled out in the bottom panel with small squares, with the same color code. In the scaled time $s$, the natural relaxation time here is  $s_R=t_R/\mu=400$.}
    }
    \label{fig:Wtotal_xf_opt_inside}
\end{figure}

The above discussion entails that, for initial points inside the limit cycle, $b\le |x_{10}| \le x_{\LC}^{\max}$, we expect the optimal final point to vary with $s_f$ from $x_{1f}^{\opt}=x_{10}$ for $s_f\ll 1$ to $x_{1f}^{\opt}=b$ for $\mu s_f\gg 1$.
In order to check this theoretical prediction, we have employed the particular case of the van der Pol oscillator. In Fig.~\ref{fig:Wtotal_xf_opt_inside}, we plot the total work as a function of the final point $x_{1f}$ (top panel) for different values of the connection time $s_f$, and the corresponding optimal value $x_{1f}^{\opt}$ as a function of $s_f$ (bottom panel). Specifically, we have taken $\mu=0.1$ and $x_{10}=1.5$. For small values of $s_f$, the dominance of $W_{\nc}$ over $\Delta E$ implies that $x_{1f}^{\opt}\simeq  x_{10}$. As $s_f$ increases, $x_{1f}^{\opt}$ starts to decrease until $x_{1f}^{\opt}\to 1$ for $s_f\to\infty$; recall that $b=1$ for the van der Pol oscillator. We have numerically checked that  $x_{1f}^{\opt}$ is  accurately determined by condition \ref{item2-lienard}, i.e. Eq.~\eqref{eq:WtotalTransVer2}, for all $s_f$.

\begin{figure}
    \centering
    \includegraphics[width=3in]{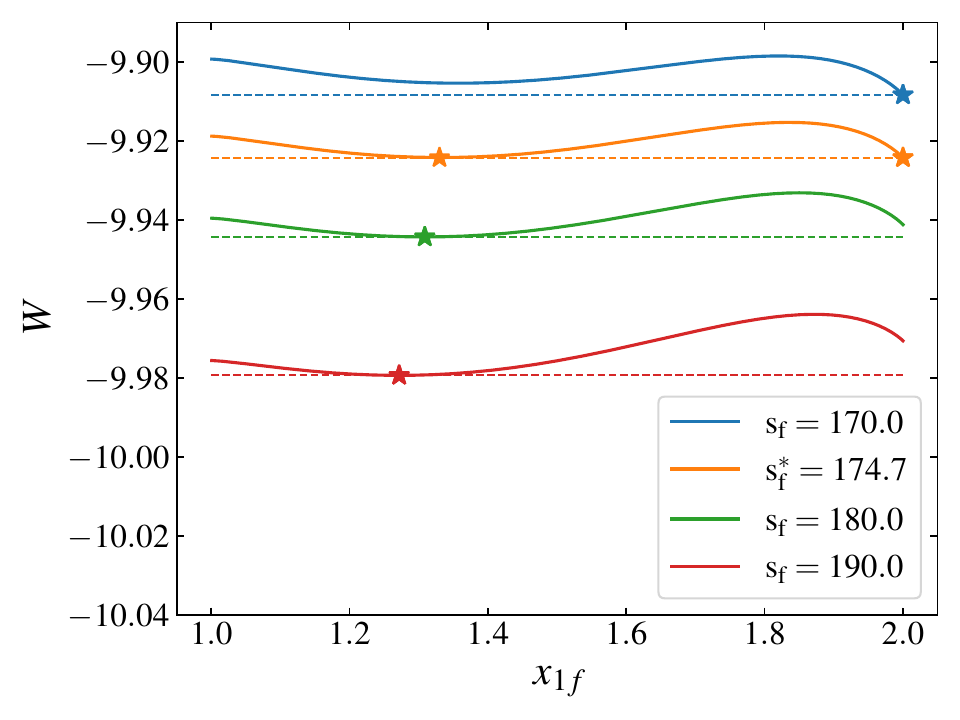}
    \includegraphics[width=3in]{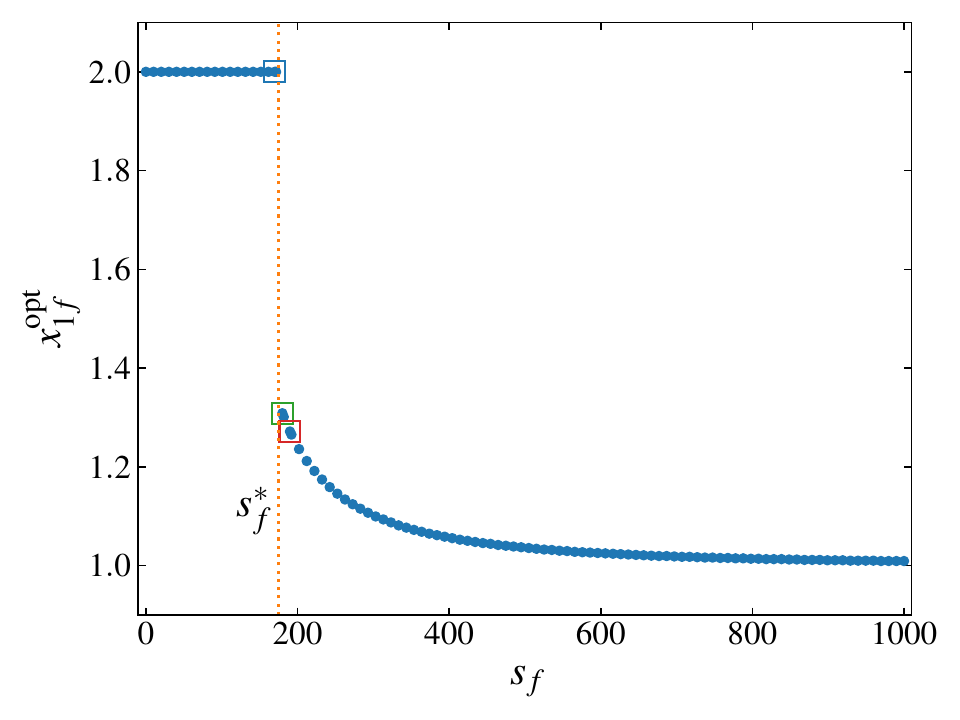}
    \caption{{Total work $W$ as a function of the final point over the limit cycle (top) and optimal final limit cycle point $x_{1f}^{\opt}$ as a function of the scaled time $s_f$. Again, the plotted curves correspond to the van der Pol oscillator with a damping constant $\mu=0.1$, but the initial point lies outside the limit cycle, namely $x_{10}=5$.  The code for the stars (top) and the small squares (bottom) is analogous to that in Fig.~\ref{fig:Wtotal_xf_opt_inside}. At $s_f=s_f^*\simeq 174.7$, it can be observed  how the local minimum of $W$ equals its value at the boundary $x_{1f}=x_{\LC}^{\max}$.}}
    \label{fig:Wtotal_xf_opt_outside}
\end{figure}

We find a  different, more complex, situation when the initial point lies outside the limit cycle, $|x_{10}|>x_{\LC}^{\max}$. Figure~\ref{fig:Wtotal_xf_opt_outside} shows the results for the minimisation of the total work, also for the van der Pol oscillator with $\mu=0.1$ as in Fig.~\ref{fig:Wtotal_xf_opt_inside} but for $x_{10}=5$. Again, for small values of $s_f$, $W$ is dominated by $W_{\nc}$. This entails that the optimal final point is given by condition \ref{item1-lienard}, i.e. $x_{1f}^{\opt} = x_{\LC}^{\max}$. This situation is maintained until a certain critical value of the connection time, $s_f=s_f^*$, is reached. Thereat, $x_{1f}^{\opt}$ abruptly changes, and from that moment on is given by condition \ref{item2-lienard}, i.e.~Eq.~\eqref{eq:WtotalTransVer2}. This sudden change can be explained taking a look at the top plot of Fig.~\ref{fig:Wtotal_xf_opt_outside}. When $s_f$ is long enough, $W$ has a local minimum given by Eq.~\eqref{eq:WtotalTransVer2}. This local minimum may be the global one depending on the value of $W$ at its boundary $x_{1f} = x_{\LC}^{\max}$. There is a critical value of the connection time $s_f=s_f^*$ where both values equate. For shorter times, $x_{1f}^{\opt}=x_{\LC}^{\max}$, whereas for longer times the optimal point for synchronisation is provided by Eq.~\eqref{eq:WtotalTransVer2}. Note that for $s_f>s_f^*$, the behaviour is identical to the previous case, $x_{1f}^{\opt}$ asymptotically tends to $b=1$ in the limit as $s_f\to\infty$.

\section{Conclusions and perspectives}\label{sec:conclusions}

In this work, we have investigated the optimal synchronisation of the van der Pol oscillator to its limit cycle. We have understood optimality in terms of the minimisation of the non-conservative work from a given initial point $\bm{x}_0=(x_{10},x_{20})$ on the phase plane to any final point $\bm{x}_f=(x_{1f},x_{2f})$ belonging to the limit cycle, in a given connection time $t_f$. The non-conservative work can be minimised if both the initial and final points lie on the same region of the phase plane where the non-conservative force is dissipative, i.e. $|x_1|\ge 1$. {Interestingly, the minimum non-conservative work $W_{\nc}^{\min}$ depends on the initial position $x_{10}$ but not on the initial velocity $x_{20}$.}

The minimisation of the non-conservative work has some interesting physical consequences. For initial points such that its initial position $x_{10}$ lies ``inside'' the limit cycle, i.e. $|x_{10}|\le x_{\LC}^{\max}$, the minimum conservative work corresponds to a vertical trajectory in phase space with constant position, $x_1(t)=x_{10}$, and zero velocity, $x_2(t)=0$, and thus zero non-conservative work. The final point over the limit cycle is reached by introducing a delta-peak force, i.e. an impulsive force, that instantaneously changes the velocity without altering the position. For initial points such that its initial position $x_{10}$ lies ``outside'' the limit cycle, i.e. $|x_{10}|> x_{\LC}^{\max}$, the optimal phase space trajectory joins the initial point with the rightmost or leftmost point of the limit cycle $(\pm x_{\LC}^{\max},0)$, depending on the sign of $x_{10}$. Analogously to recent results for the irreversible work in the field of stochastic thermodynamics, the minimum non-conservative work in this case is inversely proportional to the connection time. This implies that there appears a speed limit inequality for the synchronisation to the limit cycle.

Especially interesting is the small damping limit $\mu\ll 1$, in which the natural relaxation time $t_R$ to the limit cycle is very long, $t_R=O(\mu^{-1})$.  Therein, the speed-limit inequality reads $s_f W_{\nc}\ge I_0$, where $I_0$ is an order of unity bound---which only depends on the initial point---and $s_f=t_f/\mu$ is a scaled connection time. This speed limit inequality entails that a SST transformation with a very short connection time, $s_f=t_f/\mu=O(1)$ can be done with finite cost. Note that the acceleration obtained in this case is enormous, $t_f/t_R=O(\mu^2)$.

It is remarkable that most of the results for the van der Pol oscillator extend to the Liénard equation, including the $t_f^{-1}$ dependence of the non-conservative work and thus the emergence of a speed limit inequality. It is interesting also that the optimality of the rightmost/leftmost points of the limit cycle---in terms of minimising the non-conservative work for the synchronisation thereto---can also be established for the Liénard equation. This property follows from purely geometric arguments, stemming from the transversality condition, without needing to have a closed expression for the limit cycle.

{
We have also considered the problem of minimising the total work done by the external force, including both the conservative and non-conservative contributions thereto. From the point of view of SST, the most relevant regime is that of small damping $\mu\ll 1$ with order of unity scaled connection times $s_f$---which, as stated above, imply an enormous acceleration of the dynamics, the acceleration factor diverges as $\mu^{-2}$. Therein, the minimisation of the total work $W$ is dominated by the non-conservative contribution $W_{\nc}$, and therefore the optimal final point over the limit cycle coincides---to the lowest order---with that obtained for $W_{\nc}$. As the connection time $s_f$ increases, the non-conservative contribution to the work decreases and there appears a competition with the conservative contribution---which is expressed mathematically by the modified transversality condition \ref{item2-lienard}, Eq.~\eqref{eq:WtotalTransVer2}. For initial points inside the limit cycle, $|x_{10}|\le x_{\LC}^{\max}$, the optimal final point smoothly varies from $x_{1f}^{\opt}=x_{10}$ for $s_f\ll 1$ to $x_{1f}=1$ for $\mu s_f\gg 1$. For initial points outside the limit cycle, $|x_{10}|> x_{\LC}^{\max}$, the behaviour of the optimal final point is more complex: there appears a critical value $s_f^*$, at which $x_{1f}^{\opt}$ presents a jump discontinuity but the minimum work is continuous. Therefore, this is analogous to a first-order phase transition, with the optimal final point $x_{1f}^{\opt}$ playing the role of a order parameter. For $s_f< s_f^*$, $x_{1f}^{\opt}=x_{\LC}^{\max}$, as was the case for the non-conservative work, whereas $x_{1f}^{\opt}$ follows Eq.~\eqref{eq:WtotalTransVer2} for $s_f>s_f^*$.

Non-linear systems typically call for specific approaches, almost always involving approximations---e.g.~there is not a general theory for solving non-linear differential equations, in contrast with the situation for the linear case. Nevertheless, we have been able here to derive a whole framework for the minimisation of the work done in the synchronisation to a limit cycle in a general class of non-linear systems, with many exact analytical  results---without knowing the explicit function that gives the shape of the limit cycle. This is a remarkable result, which may pave the way to find the optimal synchronisation to self-sustained oscillations in more complex non-linear systems, which are relevant in many fields like active matter and other biological contexts.\cite{bruckner_stochastic_2019,alicki_leaking_2021,kotwal_active_2021,liu_viscoelastic_2021,hu_phototunable_2021,tucci_modeling_2022,landman_transcription_2023,zheng_self-oscillation_2023} 

{ From a practical point of view, there is a maximum value of the force that can be applied, i.e. $|F(t)|\le K$, in any actual physical system. Moreover, the force can only be applied during a non-vanishing time interval. Hence, it would be interesting to address the mimimisation of the work with the non-holonomic constraint $|F(t)|\le K$---which makes it necessary to resort to the tools of optimal control theory, specifically Pontryagin's maximum principle.\cite{pontryagin_mathematical_1987,liberzon_calculus_2012} For large enough $K$, physical intuition makes one expect a regularised version of Eq.~\eqref{F-opt-func-t-EL-with-deltas} as the solution of the optimal control, with (i) the initial and final delta peaks being substituted with very short---as compared with the smallest intrinsic timescale of the undriven system---time windows where $F(t)=\pm K$, and (ii) an intermediate time window, between the short initial and final ones, in which the solution follows the Euler-Lagrange equations.\footnote{In fact, this is the case for other underdamped control problems, see for instance Ref.~\cite{baldovin_levitated_2024}.} The expectation of the three-stage picture just described is based on the position of the particle being changed by a tiny, infinitesimal from a mathematical point of view, amount in the initial and final time windows---i.e. for large enough $K$, the initial and final $\pm K$ forces are impulsive within a high degree of approximation. As $K$ is decreased, the duration of the necessary initial and final time windows increase and become comparable to the intrinsic timescales, making the position appreciably change therein. As a consequence, the validity of the simple three-stage picture outlined above cannot be guaranteed as $K$ decreases, and a thorough analysis of the constrained control problem becomes compulsory.
}

\appendix

\section{Transversality Condition}\label{app-transvers-cond}
{

In this appendix we look into a detailed derivation of the transversality condition in a general case. Let $\bm{x}:[t_0,t_f]\to \mathbb{R}^n$ be a $\mathcal{C}^1(t_0,t_f)$ function representing the phase space trajectory of a system. Now let us consider the ``cost'' functional $S[\bm{x}]$ of a given trajectory $\bm{x}$. We assume this cost to have the general form
\begin{equation}
    S[\bm{x}] = \int_{t_0}^{t_f}dt\, L(\bm{x},\dot{\bm{x}}) + M(\bm{x}_f),
\end{equation}
i.e. $L(\bm{x},\dot{\bm{x}})$ is the ``Lagrangian'' of the problem, the time integral of which provides the running cost during the interval $(0,t_f)$, whereas $M(\bm{x}_f)$ is a $\mathcal{C}^1$ function giving the terminal cost of the trajectory, wity $\bm{x}_f \equiv \bm{x}(t_f)$. 

Now, we consider the minimisation problem
\begin{equation}
    \min_{\bm{x}\in\mathcal{C}^1(t_0,t_f)} S[\bm{x}],
\end{equation}
which is known in optimal control theory as the Bolza problem~\cite{liberzon_calculus_2012}. For our purposes, we analyse the Bolza problem subject to the boundary conditions 
\begin{equation}
    \bm{x}(t_0)= \bm{x}_0, \quad g(\bm{x}_f)=0,
\end{equation}
where $g$ is a given function. That is, the initial point is fixed, while the final point lies on a certain target set $g(\bm{x}_f)=0$.

The first-order necessary condition for optimality is $\delta S=0$. Since $\bm{x}_0$ is fixed, $\delta \bm{x}_0=0$. This means
\begin{align}
    \delta S =& \int_{t_0}^{t_f} dt \, \left[ \delta \bm{x} \cdot \nabla_{\bm{x}} L(\bm{x},\dot{\bm{x}}) + \delta \dot{\bm{x}} \cdot \nabla_{\dot{\bm{x}}} L(\bm{x},\dot{\bm{x}})\right]\nonumber \\  &+  \delta \bm{x}_f \cdot \nabla_{\bm{x}_f} M(\bm{x}_f) = 0.
\end{align}
Integrating by parts the second term of the integral, we get
\begin{align}\label{eq:deltaS-v1}
    \delta S =& \int_{t_0}^{t_f} dt \, \delta \bm{x} \cdot \left[\nabla_{\bm{x}} L(\bm{x},\dot{\bm{x}}) - \frac{d}{dt}\nabla_{\dot{\bm{x}}} L(\bm{x},\dot{\bm{x}})\right]  \nonumber \\ &+ \delta \bm{x}_f \cdot \left[ \bm{p}_f+\nabla_{\bm{x}_f} M(\bm{x}_f) \right]= 0,
\end{align}
where we have defined the momenta and its final value as
\begin{equation}
    \bm{p}\equiv \nabla_{\dot{\bm{x}}}L(\bm{x},\dot{\bm{x}}), \quad \bm{p}_f\equiv \bm{p}(t_f).
\end{equation}
If $\bm{x}_f$ were fixed---as is the case in the least action principle of classical mechanics~\cite{landau_mechanics_1976}, the last term of Eq.~\eqref{eq:deltaS-v1} would identically vanish. Then, the arbitrariness of $\delta\bm{x}(t)$ in the time interval $(0,t_f)$ leads to the Euler-Lagrange equations
\begin{equation}
    \nabla_{\bm{x}} L(\bm{x},\dot{\bm{x}}) - \frac{d}{dt}\nabla_{\dot{\bm{x}}} L(\bm{x},\dot{\bm{x}})=\bm{0}.
\end{equation}
In our case, $\bm{x}_f$ is not fixed, it lies on the curve $g(\bm{x}_f)=0$. Yet,  since $\delta\bm{x}(t)$ for $t\in(0,t_f)$ and $\delta\bm{x}_f$ can be independently varied, (i) the Euler-Lagrange equations continue to hold and (ii) the boundary term in Eq.~\eqref{eq:deltaS-v1} must also vanish, i.e.
\begin{equation}
    \delta \bm{x}_f \cdot \left[ \bm{p}_f+\nabla_{\bm{x}_f} M(\bm{x}_f) \right]=0.
\end{equation}
This condition is known as the transversality condition, since it tells us that $\bm{p}_f+\nabla_{\bm{x}_f} M(\bm{x}_f)$ must be orthogonal to $\delta\bm{x}_f$. It selects, among all the points on the target set $g(\bm{x}_f)=0$, the optimal final point. 

In the main text, we have first considered the problem of the minimisation of the non-conservative work. In that case, there is no terminal cost and the transversality condition reduces to $\delta\bm{x}_f\cdot\bm{p}_f=0$, Eq.~\eqref{eq:transvers-cond}. Later, we have considered the problem of minimising the total work. The inclusion of the conservative contribution thereto entails that there is now a terminal cost $E(\bm{x}_f)$, as expressed by Eq.~\eqref{eq:total-work}. The transversality condition thus changes to $\delta\bm{x}_f\cdot\left[\bm{p}_f+\nabla_{\bm{x}_f} E(\bm{x}_f)\right]=0$, Eq.~\eqref{eq:extended-transvers}. It is to differentiate Eq.~\eqref{eq:extended-transvers} from Eq.~\eqref{eq:transvers-cond} that we have referred to the former as the modified transversality condition in this paper.
}


\begin{acknowledgments}
CRM, CAP and AP acknowledge financial support from Grant
  PID2021-122588NB-I00 funded by MCIN/AEI/10.13039/501100011033/ and
  by ``ERDF A way of making Europe''. All authors
  acknowledge financial support from Grant ProyExcel\_00796 funded by
  Junta de Andalucía's PAIDI 2020 program. CAP acknowledges the
  funding received from European Union’s Horizon Europe–Marie
  Skłodowska-Curie 2021 program through the Postdoctoral Fellowship
  with Reference 101065902 (ORION). CRM acknowledges support from Ministry of Science, Innovation and Universities FPU programme through Grant FPU22/01151.
\end{acknowledgments}

\section*{Data Availability Statement}

 The Python codes employed for generating the data and figures that support the findings of this study are openly available in the~\href{https://github.com/fine-group-us/Optimal-synchronisation-to-a-limit-cycle}{GitHub page} of University of Sevilla's FINE research group.

\bibliography{Mi-biblioteca-13-jun-2024}

\begin{thebibliography}{66}%
\makeatletter
\providecommand \@ifxundefined [1]{%
 \@ifx{#1\undefined}
}%
\providecommand \@ifnum [1]{%
 \ifnum #1\expandafter \@firstoftwo
 \else \expandafter \@secondoftwo
 \fi
}%
\providecommand \@ifx [1]{%
 \ifx #1\expandafter \@firstoftwo
 \else \expandafter \@secondoftwo
 \fi
}%
\providecommand \natexlab [1]{#1}%
\providecommand \enquote  [1]{``#1''}%
\providecommand \bibnamefont  [1]{#1}%
\providecommand \bibfnamefont [1]{#1}%
\providecommand \citenamefont [1]{#1}%
\providecommand \href@noop [0]{\@secondoftwo}%
\providecommand \href [0]{\begingroup \@sanitize@url \@href}%
\providecommand \@href[1]{\@@startlink{#1}\@@href}%
\providecommand \@@href[1]{\endgroup#1\@@endlink}%
\providecommand \@sanitize@url [0]{\catcode `\\12\catcode `\$12\catcode
  `\&12\catcode `\#12\catcode `\^12\catcode `\_12\catcode `\%12\relax}%
\providecommand \@@startlink[1]{}%
\providecommand \@@endlink[0]{}%
\providecommand \url  [0]{\begingroup\@sanitize@url \@url }%
\providecommand \@url [1]{\endgroup\@href {#1}{\urlprefix }}%
\providecommand \urlprefix  [0]{URL }%
\providecommand \Eprint [0]{\href }%
\providecommand \doibase [0]{http://dx.doi.org/}%
\providecommand \selectlanguage [0]{\@gobble}%
\providecommand \bibinfo  [0]{\@secondoftwo}%
\providecommand \bibfield  [0]{\@secondoftwo}%
\providecommand \translation [1]{[#1]}%
\providecommand \BibitemOpen [0]{}%
\providecommand \bibitemStop [0]{}%
\providecommand \bibitemNoStop [0]{.\EOS\space}%
\providecommand \EOS [0]{\spacefactor3000\relax}%
\providecommand \BibitemShut  [1]{\csname bibitem#1\endcsname}%
\let\auto@bib@innerbib\@empty
\bibitem [{\citenamefont {Ginoux}\ and\ \citenamefont
  {Letellier}(2012)}]{ginoux_van_2012}%
  \BibitemOpen
  \bibfield  {author} {\bibinfo {author} {\bibfnamefont {J.-M.}\ \bibnamefont
  {Ginoux}}\ and\ \bibinfo {author} {\bibfnamefont {C.}~\bibnamefont
  {Letellier}},\ }\bibfield  {title} {\enquote {\bibinfo {title} {Van der {Pol}
  and the history of relaxation oscillations: {Toward} the emergence of a
  concept},}\ }\href {\doibase 10.1063/1.3670008} {\bibfield  {journal}
  {\bibinfo  {journal} {Chaos: An Interdisciplinary Journal of Nonlinear
  Science}\ }\textbf {\bibinfo {volume} {22}},\ \bibinfo {pages} {023120}
  (\bibinfo {year} {2012})}\BibitemShut {NoStop}%
\bibitem [{\citenamefont {Jenkins}(2013)}]{jenkins_self-oscillation_2013}%
  \BibitemOpen
  \bibfield  {author} {\bibinfo {author} {\bibfnamefont {A.}~\bibnamefont
  {Jenkins}},\ }\bibfield  {title} {\enquote {\bibinfo {title}
  {Self-oscillation},}\ }\href {\doibase 10.1016/j.physrep.2012.10.007}
  {\bibfield  {journal} {\bibinfo  {journal} {Physics Reports}\ }\textbf
  {\bibinfo {volume} {525}},\ \bibinfo {pages} {167--222} (\bibinfo {year}
  {2013})}\BibitemShut {NoStop}%
\bibitem [{\citenamefont {Guéry-Odelin}\ \emph {et~al.}(2019)\citenamefont
  {Guéry-Odelin}, \citenamefont {Ruschhaupt}, \citenamefont {Kiely},
  \citenamefont {Torrontegui}, \citenamefont {Martínez-Garaot},\ and\
  \citenamefont {Muga}}]{guery-odelin_shortcuts_2019}%
  \BibitemOpen
  \bibfield  {author} {\bibinfo {author} {\bibfnamefont {D.}~\bibnamefont
  {Guéry-Odelin}}, \bibinfo {author} {\bibfnamefont {A.}~\bibnamefont
  {Ruschhaupt}}, \bibinfo {author} {\bibfnamefont {A.}~\bibnamefont {Kiely}},
  \bibinfo {author} {\bibfnamefont {E.}~\bibnamefont {Torrontegui}}, \bibinfo
  {author} {\bibfnamefont {S.}~\bibnamefont {Martínez-Garaot}}, \ and\
  \bibinfo {author} {\bibfnamefont {J.~G.}\ \bibnamefont {Muga}},\ }\bibfield
  {title} {\enquote {\bibinfo {title} {Shortcuts to adiabaticity: {Concepts},
  methods, and applications},}\ }\href {\doibase 10.1103/RevModPhys.91.045001}
  {\bibfield  {journal} {\bibinfo  {journal} {Reviews of Modern Physics}\
  }\textbf {\bibinfo {volume} {91}},\ \bibinfo {pages} {045001} (\bibinfo
  {year} {2019})}\BibitemShut {NoStop}%
\bibitem [{\citenamefont {Guéry-Odelin}\ \emph {et~al.}(2023)\citenamefont
  {Guéry-Odelin}, \citenamefont {Jarzynski}, \citenamefont {Plata},
  \citenamefont {Prados},\ and\ \citenamefont
  {Trizac}}]{guery-odelin_driving_2023}%
  \BibitemOpen
  \bibfield  {author} {\bibinfo {author} {\bibfnamefont {D.}~\bibnamefont
  {Guéry-Odelin}}, \bibinfo {author} {\bibfnamefont {C.}~\bibnamefont
  {Jarzynski}}, \bibinfo {author} {\bibfnamefont {C.~A.}\ \bibnamefont
  {Plata}}, \bibinfo {author} {\bibfnamefont {A.}~\bibnamefont {Prados}}, \
  and\ \bibinfo {author} {\bibfnamefont {E.}~\bibnamefont {Trizac}},\
  }\bibfield  {title} {\enquote {\bibinfo {title} {Driving rapidly while
  remaining in control: classical shortcuts from {Hamiltonian} to stochastic
  dynamics},}\ }\href {\doibase 10.1088/1361-6633/acacad} {\bibfield  {journal}
  {\bibinfo  {journal} {Reports on Progress in Physics}\ }\textbf {\bibinfo
  {volume} {86}},\ \bibinfo {pages} {035902} (\bibinfo {year}
  {2023})}\BibitemShut {NoStop}%
\bibitem [{\citenamefont {Van Der~Pol}(1926)}]{van_der_pol_lxxxviii_1926}%
  \BibitemOpen
  \bibfield  {author} {\bibinfo {author} {\bibfnamefont {B.}~\bibnamefont {Van
  Der~Pol}},\ }\bibfield  {title} {\enquote {\bibinfo {title} {{LXXXVIII}.
  \textit{{On} “relaxation-oscillations”}},}\ }\href {\doibase
  10.1080/14786442608564127} {\bibfield  {journal} {\bibinfo  {journal} {The
  London, Edinburgh, and Dublin Philosophical Magazine and Journal of Science}\
  }\textbf {\bibinfo {volume} {2}},\ \bibinfo {pages} {978--992} (\bibinfo
  {year} {1926})}\BibitemShut {NoStop}%
\bibitem [{\citenamefont {Sakaguchi}(2009)}]{sakaguchi_limit-cycle_2009}%
  \BibitemOpen
  \bibfield  {author} {\bibinfo {author} {\bibfnamefont {H.}~\bibnamefont
  {Sakaguchi}},\ }\bibfield  {title} {\enquote {\bibinfo {title} {Limit-cycle
  oscillation of an elastic filament and caterpillar motion},}\ }\href
  {\doibase 10.1103/PhysRevE.79.026216} {\bibfield  {journal} {\bibinfo
  {journal} {Physical Review E}\ }\textbf {\bibinfo {volume} {79}},\ \bibinfo
  {pages} {026216} (\bibinfo {year} {2009})}\BibitemShut {NoStop}%
\bibitem [{\citenamefont {Alicki}, \citenamefont {Gelbwaser-Klimovsky},\ and\
  \citenamefont {Jenkins}(2021)}]{alicki_leaking_2021}%
  \BibitemOpen
  \bibfield  {author} {\bibinfo {author} {\bibfnamefont {R.}~\bibnamefont
  {Alicki}}, \bibinfo {author} {\bibfnamefont {D.}~\bibnamefont
  {Gelbwaser-Klimovsky}}, \ and\ \bibinfo {author} {\bibfnamefont
  {A.}~\bibnamefont {Jenkins}},\ }\bibfield  {title} {\enquote {\bibinfo
  {title} {Leaking elastic capacitor as model for active matter},}\ }\href
  {\doibase 10.1103/PhysRevE.103.052131} {\bibfield  {journal} {\bibinfo
  {journal} {Physical Review E}\ }\textbf {\bibinfo {volume} {103}},\ \bibinfo
  {pages} {052131} (\bibinfo {year} {2021})}\BibitemShut {NoStop}%
\bibitem [{\citenamefont {Kotwal}\ \emph {et~al.}(2021)\citenamefont {Kotwal},
  \citenamefont {Moseley}, \citenamefont {Stegmaier}, \citenamefont {Imhof},
  \citenamefont {Brand}, \citenamefont {Kießling}, \citenamefont {Thomale},
  \citenamefont {Ronellenfitsch},\ and\ \citenamefont
  {Dunkel}}]{kotwal_active_2021}%
  \BibitemOpen
  \bibfield  {author} {\bibinfo {author} {\bibfnamefont {T.}~\bibnamefont
  {Kotwal}}, \bibinfo {author} {\bibfnamefont {F.}~\bibnamefont {Moseley}},
  \bibinfo {author} {\bibfnamefont {A.}~\bibnamefont {Stegmaier}}, \bibinfo
  {author} {\bibfnamefont {S.}~\bibnamefont {Imhof}}, \bibinfo {author}
  {\bibfnamefont {H.}~\bibnamefont {Brand}}, \bibinfo {author} {\bibfnamefont
  {T.}~\bibnamefont {Kießling}}, \bibinfo {author} {\bibfnamefont
  {R.}~\bibnamefont {Thomale}}, \bibinfo {author} {\bibfnamefont
  {H.}~\bibnamefont {Ronellenfitsch}}, \ and\ \bibinfo {author} {\bibfnamefont
  {J.}~\bibnamefont {Dunkel}},\ }\bibfield  {title} {\enquote {\bibinfo {title}
  {Active topolectrical circuits},}\ }\href {\doibase 10.1073/pnas.2106411118}
  {\bibfield  {journal} {\bibinfo  {journal} {Proceedings of the National
  Academy of Sciences}\ }\textbf {\bibinfo {volume} {118}},\ \bibinfo {pages}
  {e2106411118} (\bibinfo {year} {2021})}\BibitemShut {NoStop}%
\bibitem [{\citenamefont {Liu}\ \emph {et~al.}(2021)\citenamefont {Liu},
  \citenamefont {Shankar}, \citenamefont {Marchetti},\ and\ \citenamefont
  {Wu}}]{liu_viscoelastic_2021}%
  \BibitemOpen
  \bibfield  {author} {\bibinfo {author} {\bibfnamefont {S.}~\bibnamefont
  {Liu}}, \bibinfo {author} {\bibfnamefont {S.}~\bibnamefont {Shankar}},
  \bibinfo {author} {\bibfnamefont {M.~C.}\ \bibnamefont {Marchetti}}, \ and\
  \bibinfo {author} {\bibfnamefont {Y.}~\bibnamefont {Wu}},\ }\bibfield
  {title} {\enquote {\bibinfo {title} {Viscoelastic control of spatiotemporal
  order in bacterial active matter},}\ }\href {\doibase
  10.1038/s41586-020-03168-6} {\bibfield  {journal} {\bibinfo  {journal}
  {Nature}\ }\textbf {\bibinfo {volume} {590}},\ \bibinfo {pages} {80--84}
  (\bibinfo {year} {2021})}\BibitemShut {NoStop}%
\bibitem [{\citenamefont {Tucci}\ \emph {et~al.}(2022)\citenamefont {Tucci},
  \citenamefont {Rold\'an}, \citenamefont {Gambassi}, \citenamefont {Belousov},
  \citenamefont {Berger}, \citenamefont {Alonso},\ and\ \citenamefont
  {Hudspeth}}]{tucci_modeling_2022}%
  \BibitemOpen
  \bibfield  {author} {\bibinfo {author} {\bibfnamefont {G.}~\bibnamefont
  {Tucci}}, \bibinfo {author} {\bibfnamefont {E.}~\bibnamefont {Rold\'an}},
  \bibinfo {author} {\bibfnamefont {A.}~\bibnamefont {Gambassi}}, \bibinfo
  {author} {\bibfnamefont {R.}~\bibnamefont {Belousov}}, \bibinfo {author}
  {\bibfnamefont {F.}~\bibnamefont {Berger}}, \bibinfo {author} {\bibfnamefont
  {R.}~\bibnamefont {Alonso}}, \ and\ \bibinfo {author} {\bibfnamefont
  {A.}~\bibnamefont {Hudspeth}},\ }\bibfield  {title} {\enquote {\bibinfo
  {title} {Modeling {Active} {Non}-{Markovian} {Oscillations}},}\ }\href
  {\doibase 10.1103/PhysRevLett.129.030603} {\bibfield  {journal} {\bibinfo
  {journal} {Physical Review Letters}\ }\textbf {\bibinfo {volume} {129}},\
  \bibinfo {pages} {030603} (\bibinfo {year} {2022})}\BibitemShut {NoStop}%
\bibitem [{\citenamefont {Leloup}, \citenamefont {Gonze},\ and\ \citenamefont
  {Goldbeter}(1999)}]{leloup_limit_1999}%
  \BibitemOpen
  \bibfield  {author} {\bibinfo {author} {\bibfnamefont {J.-C.}\ \bibnamefont
  {Leloup}}, \bibinfo {author} {\bibfnamefont {D.}~\bibnamefont {Gonze}}, \
  and\ \bibinfo {author} {\bibfnamefont {A.}~\bibnamefont {Goldbeter}},\
  }\bibfield  {title} {\enquote {\bibinfo {title} {Limit {Cycle} {Models} for
  {Circadian} {Rhythms} {Based} on {Transcriptional} {Regulation} in
  \textit{{Drosophila}} and \textit{{Neurospora}}},}\ }\href {\doibase
  10.1177/074873099129000948} {\bibfield  {journal} {\bibinfo  {journal}
  {Journal of Biological Rhythms}\ }\textbf {\bibinfo {volume} {14}},\ \bibinfo
  {pages} {433--448} (\bibinfo {year} {1999})}\BibitemShut {NoStop}%
\bibitem [{\citenamefont {Roenneberg}\ \emph {et~al.}(2008)\citenamefont
  {Roenneberg}, \citenamefont {Chua}, \citenamefont {Bernardo},\ and\
  \citenamefont {Mendoza}}]{roenneberg_modelling_2008}%
  \BibitemOpen
  \bibfield  {author} {\bibinfo {author} {\bibfnamefont {T.}~\bibnamefont
  {Roenneberg}}, \bibinfo {author} {\bibfnamefont {E.~J.}\ \bibnamefont
  {Chua}}, \bibinfo {author} {\bibfnamefont {R.}~\bibnamefont {Bernardo}}, \
  and\ \bibinfo {author} {\bibfnamefont {E.}~\bibnamefont {Mendoza}},\
  }\bibfield  {title} {\enquote {\bibinfo {title} {Modelling {Biological}
  {Rhythms}},}\ }\href {\doibase 10.1016/j.cub.2008.07.017} {\bibfield
  {journal} {\bibinfo  {journal} {Current Biology}\ }\textbf {\bibinfo {volume}
  {18}},\ \bibinfo {pages} {R826--R835} (\bibinfo {year} {2008})}\BibitemShut
  {NoStop}%
\bibitem [{\citenamefont {Brückner}\ \emph {et~al.}(2019)\citenamefont
  {Brückner}, \citenamefont {Fink}, \citenamefont {Schreiber}, \citenamefont
  {Röttgermann}, \citenamefont {Rädler},\ and\ \citenamefont
  {Broedersz}}]{bruckner_stochastic_2019}%
  \BibitemOpen
  \bibfield  {author} {\bibinfo {author} {\bibfnamefont {D.~B.}\ \bibnamefont
  {Brückner}}, \bibinfo {author} {\bibfnamefont {A.}~\bibnamefont {Fink}},
  \bibinfo {author} {\bibfnamefont {C.}~\bibnamefont {Schreiber}}, \bibinfo
  {author} {\bibfnamefont {P.~J.~F.}\ \bibnamefont {Röttgermann}}, \bibinfo
  {author} {\bibfnamefont {J.~O.}\ \bibnamefont {Rädler}}, \ and\ \bibinfo
  {author} {\bibfnamefont {C.~P.}\ \bibnamefont {Broedersz}},\ }\bibfield
  {title} {\enquote {\bibinfo {title} {Stochastic nonlinear dynamics of
  confined cell migration in two-state systems},}\ }\href {\doibase
  10.1038/s41567-019-0445-4} {\bibfield  {journal} {\bibinfo  {journal} {Nature
  Physics}\ }\textbf {\bibinfo {volume} {15}},\ \bibinfo {pages} {595--601}
  (\bibinfo {year} {2019})}\BibitemShut {NoStop}%
\bibitem [{\citenamefont {FitzHugh}(1961)}]{fitzhugh_impulses_1961}%
  \BibitemOpen
  \bibfield  {author} {\bibinfo {author} {\bibfnamefont {R.}~\bibnamefont
  {FitzHugh}},\ }\bibfield  {title} {\enquote {\bibinfo {title} {Impulses and
  {Physiological} {States} in {Theoretical} {Models} of {Nerve} {Membrane}},}\
  }\href {\doibase 10.1016/S0006-3495(61)86902-6} {\bibfield  {journal}
  {\bibinfo  {journal} {Biophysical Journal}\ }\textbf {\bibinfo {volume}
  {1}},\ \bibinfo {pages} {445--466} (\bibinfo {year} {1961})}\BibitemShut
  {NoStop}%
\bibitem [{\citenamefont {Nagumo}, \citenamefont {Arimoto},\ and\ \citenamefont
  {Yoshizawa}(1962)}]{nagumo_active_1962}%
  \BibitemOpen
  \bibfield  {author} {\bibinfo {author} {\bibfnamefont {J.}~\bibnamefont
  {Nagumo}}, \bibinfo {author} {\bibfnamefont {S.}~\bibnamefont {Arimoto}}, \
  and\ \bibinfo {author} {\bibfnamefont {S.}~\bibnamefont {Yoshizawa}},\
  }\bibfield  {title} {\enquote {\bibinfo {title} {An {Active} {Pulse}
  {Transmission} {Line} {Simulating} {Nerve} {Axon}},}\ }\href {\doibase
  10.1109/JRPROC.1962.288235} {\bibfield  {journal} {\bibinfo  {journal}
  {Proceedings of the IRE}\ }\textbf {\bibinfo {volume} {50}},\ \bibinfo
  {pages} {2061--2070} (\bibinfo {year} {1962})}\BibitemShut {NoStop}%
\bibitem [{\citenamefont {Cartwright}\ \emph {et~al.}(1999)\citenamefont
  {Cartwright}, \citenamefont {Eguíluz}, \citenamefont {Hernández-García},\
  and\ \citenamefont {Piro}}]{cartwright_dynamics_1999}%
  \BibitemOpen
  \bibfield  {author} {\bibinfo {author} {\bibfnamefont {J.~H.~E.}\
  \bibnamefont {Cartwright}}, \bibinfo {author} {\bibfnamefont {V.~M.}\
  \bibnamefont {Eguíluz}}, \bibinfo {author} {\bibfnamefont {E.}~\bibnamefont
  {Hernández-García}}, \ and\ \bibinfo {author} {\bibfnamefont
  {O.}~\bibnamefont {Piro}},\ }\bibfield  {title} {\enquote {\bibinfo {title}
  {Dynamics of {Elastic} {Excitable} {Media}},}\ }\href {\doibase
  10.1142/S0218127499001620} {\bibfield  {journal} {\bibinfo  {journal}
  {International Journal of Bifurcation and Chaos}\ }\textbf {\bibinfo {volume}
  {09}},\ \bibinfo {pages} {2197--2202} (\bibinfo {year} {1999})}\BibitemShut
  {NoStop}%
\bibitem [{\citenamefont {Romanczuk}\ \emph {et~al.}(2012)\citenamefont
  {Romanczuk}, \citenamefont {Bär}, \citenamefont {Ebeling}, \citenamefont
  {Lindner},\ and\ \citenamefont {Schimansky-Geier}}]{romanczuk_active_2012}%
  \BibitemOpen
  \bibfield  {author} {\bibinfo {author} {\bibfnamefont {P.}~\bibnamefont
  {Romanczuk}}, \bibinfo {author} {\bibfnamefont {M.}~\bibnamefont {Bär}},
  \bibinfo {author} {\bibfnamefont {W.}~\bibnamefont {Ebeling}}, \bibinfo
  {author} {\bibfnamefont {B.}~\bibnamefont {Lindner}}, \ and\ \bibinfo
  {author} {\bibfnamefont {L.}~\bibnamefont {Schimansky-Geier}},\ }\bibfield
  {title} {\enquote {\bibinfo {title} {Active {Brownian} particles: {From}
  individual to collective stochastic dynamics},}\ }\href {\doibase
  10.1140/epjst/e2012-01529-y} {\bibfield  {journal} {\bibinfo  {journal} {The
  European Physical Journal Special Topics}\ }\textbf {\bibinfo {volume}
  {202}},\ \bibinfo {pages} {1--162} (\bibinfo {year} {2012})}\BibitemShut
  {NoStop}%
\bibitem [{\citenamefont {Strogatz}(2024)}]{strogatz_nonlinear_2024}%
  \BibitemOpen
  \bibfield  {author} {\bibinfo {author} {\bibfnamefont {S.~H.}\ \bibnamefont
  {Strogatz}},\ }\href {\doibase 10.1201/9780429398490} {\emph {\bibinfo
  {title} {Nonlinear {Dynamics} and {Chaos}: {With} {Applications} to
  {Physics}, {Biology}, {Chemistry}, and {Engineering}}}},\ \bibinfo {edition}
  {3rd}\ ed.\ (\bibinfo  {publisher} {Chapman and Hall/CRC},\ \bibinfo
  {address} {Boca Raton},\ \bibinfo {year} {2024})\BibitemShut {NoStop}%
\bibitem [{\citenamefont {Impens}\ and\ \citenamefont
  {Guéry-Odelin}(2023)}]{impens_shortcut_2023}%
  \BibitemOpen
  \bibfield  {author} {\bibinfo {author} {\bibfnamefont {F.}~\bibnamefont
  {Impens}}\ and\ \bibinfo {author} {\bibfnamefont {D.}~\bibnamefont
  {Guéry-Odelin}},\ }\bibfield  {title} {\enquote {\bibinfo {title} {Shortcut
  to synchronization in classical and quantum systems},}\ }\href {\doibase
  10.1038/s41598-022-27130-w} {\bibfield  {journal} {\bibinfo  {journal}
  {Scientific Reports}\ }\textbf {\bibinfo {volume} {13}},\ \bibinfo {pages}
  {453} (\bibinfo {year} {2023})}\BibitemShut {NoStop}%
\bibitem [{\citenamefont {Schmiedl}\ and\ \citenamefont
  {Seifert}(2007)}]{schmiedl_optimal_2007}%
  \BibitemOpen
  \bibfield  {author} {\bibinfo {author} {\bibfnamefont {T.}~\bibnamefont
  {Schmiedl}}\ and\ \bibinfo {author} {\bibfnamefont {U.}~\bibnamefont
  {Seifert}},\ }\bibfield  {title} {\enquote {\bibinfo {title} {Optimal
  {Finite}-{Time} {Processes} {In} {Stochastic} {Thermodynamics}},}\ }\href
  {\doibase 10.1103/PhysRevLett.98.108301} {\bibfield  {journal} {\bibinfo
  {journal} {Physical Review Letters}\ }\textbf {\bibinfo {volume} {98}},\
  \bibinfo {pages} {108301} (\bibinfo {year} {2007})}\BibitemShut {NoStop}%
\bibitem [{\citenamefont {Aurell}, \citenamefont {Mejía-Monasterio},\ and\
  \citenamefont {Muratore-Ginanneschi}(2011)}]{aurell_optimal_2011}%
  \BibitemOpen
  \bibfield  {author} {\bibinfo {author} {\bibfnamefont {E.}~\bibnamefont
  {Aurell}}, \bibinfo {author} {\bibfnamefont {C.}~\bibnamefont
  {Mejía-Monasterio}}, \ and\ \bibinfo {author} {\bibfnamefont
  {P.}~\bibnamefont {Muratore-Ginanneschi}},\ }\bibfield  {title} {\enquote
  {\bibinfo {title} {Optimal {Protocols} and {Optimal} {Transport} in
  {Stochastic} {Thermodynamics}},}\ }\href {\doibase
  10.1103/PhysRevLett.106.250601} {\bibfield  {journal} {\bibinfo  {journal}
  {Physical Review Letters}\ }\textbf {\bibinfo {volume} {106}},\ \bibinfo
  {pages} {250601} (\bibinfo {year} {2011})}\BibitemShut {NoStop}%
\bibitem [{\citenamefont {Plata}\ \emph {et~al.}(2019)\citenamefont {Plata},
  \citenamefont {Guéry-Odelin}, \citenamefont {Trizac},\ and\ \citenamefont
  {Prados}}]{plata_optimal_2019}%
  \BibitemOpen
  \bibfield  {author} {\bibinfo {author} {\bibfnamefont {C.~A.}\ \bibnamefont
  {Plata}}, \bibinfo {author} {\bibfnamefont {D.}~\bibnamefont
  {Guéry-Odelin}}, \bibinfo {author} {\bibfnamefont {E.}~\bibnamefont
  {Trizac}}, \ and\ \bibinfo {author} {\bibfnamefont {A.}~\bibnamefont
  {Prados}},\ }\bibfield  {title} {\enquote {\bibinfo {title} {Optimal work in
  a harmonic trap with bounded stiffness},}\ }\href {\doibase
  10.1103/PhysRevE.99.012140} {\bibfield  {journal} {\bibinfo  {journal}
  {Physical Review E}\ }\textbf {\bibinfo {volume} {99}},\ \bibinfo {pages}
  {012140} (\bibinfo {year} {2019})}\BibitemShut {NoStop}%
\bibitem [{\citenamefont {Zhang}(2020)}]{zhang_work_2020}%
  \BibitemOpen
  \bibfield  {author} {\bibinfo {author} {\bibfnamefont {Y.}~\bibnamefont
  {Zhang}},\ }\bibfield  {title} {\enquote {\bibinfo {title} {Work needed to
  drive a thermodynamic system between two distributions},}\ }\href {\doibase
  10.1209/0295-5075/128/30002} {\bibfield  {journal} {\bibinfo  {journal} {EPL
  (Europhysics Letters)}\ }\textbf {\bibinfo {volume} {128}},\ \bibinfo {pages}
  {30002} (\bibinfo {year} {2020})}\BibitemShut {NoStop}%
\bibitem [{\citenamefont {Sivak}\ and\ \citenamefont
  {Crooks}(2012)}]{sivak_thermodynamic_2012}%
  \BibitemOpen
  \bibfield  {author} {\bibinfo {author} {\bibfnamefont {D.~A.}\ \bibnamefont
  {Sivak}}\ and\ \bibinfo {author} {\bibfnamefont {G.~E.}\ \bibnamefont
  {Crooks}},\ }\bibfield  {title} {\enquote {\bibinfo {title} {Thermodynamic
  {Metrics} and {Optimal} {Paths}},}\ }\href {\doibase
  10.1103/PhysRevLett.108.190602} {\bibfield  {journal} {\bibinfo  {journal}
  {Physical Review Letters}\ }\textbf {\bibinfo {volume} {108}},\ \bibinfo
  {pages} {190602} (\bibinfo {year} {2012})}\BibitemShut {NoStop}%
\bibitem [{\citenamefont {Deffner}\ and\ \citenamefont
  {Campbell}(2017)}]{deffner_quantum_2017}%
  \BibitemOpen
  \bibfield  {author} {\bibinfo {author} {\bibfnamefont {S.}~\bibnamefont
  {Deffner}}\ and\ \bibinfo {author} {\bibfnamefont {S.}~\bibnamefont
  {Campbell}},\ }\bibfield  {title} {\enquote {\bibinfo {title} {Quantum speed
  limits: from {Heisenberg}’s uncertainty principle to optimal quantum
  control},}\ }\href {\doibase 10.1088/1751-8121/aa86c6} {\bibfield  {journal}
  {\bibinfo  {journal} {Journal of Physics A: Mathematical and Theoretical}\
  }\textbf {\bibinfo {volume} {50}},\ \bibinfo {pages} {453001} (\bibinfo
  {year} {2017})}\BibitemShut {NoStop}%
\bibitem [{\citenamefont {Okuyama}\ and\ \citenamefont
  {Ohzeki}(2018)}]{okuyama_quantum_2018}%
  \BibitemOpen
  \bibfield  {author} {\bibinfo {author} {\bibfnamefont {M.}~\bibnamefont
  {Okuyama}}\ and\ \bibinfo {author} {\bibfnamefont {M.}~\bibnamefont
  {Ohzeki}},\ }\bibfield  {title} {\enquote {\bibinfo {title} {Quantum {Speed}
  {Limit} is {Not} {Quantum}},}\ }\href@noop {} {\bibfield  {journal} {\bibinfo
   {journal} {Physical Review Letters}\ }\textbf {\bibinfo {volume} {120}},\
  \bibinfo {pages} {070402} (\bibinfo {year} {2018})}\BibitemShut {NoStop}%
\bibitem [{\citenamefont {Shiraishi}, \citenamefont {Funo},\ and\ \citenamefont
  {Saito}(2018)}]{shiraishi_speed_2018}%
  \BibitemOpen
  \bibfield  {author} {\bibinfo {author} {\bibfnamefont {N.}~\bibnamefont
  {Shiraishi}}, \bibinfo {author} {\bibfnamefont {K.}~\bibnamefont {Funo}}, \
  and\ \bibinfo {author} {\bibfnamefont {K.}~\bibnamefont {Saito}},\ }\bibfield
   {title} {\enquote {\bibinfo {title} {Speed {Limit} for {Classical}
  {Stochastic} {Processes}},}\ }\href@noop {} {\bibfield  {journal} {\bibinfo
  {journal} {Physical Review Letters}\ }\textbf {\bibinfo {volume} {121}},\
  \bibinfo {pages} {070601} (\bibinfo {year} {2018})}\BibitemShut {NoStop}%
\bibitem [{\citenamefont {Shanahan}\ \emph {et~al.}(2018)\citenamefont
  {Shanahan}, \citenamefont {Chenu}, \citenamefont {Margolus},\ and\
  \citenamefont {del Campo}}]{shanahan_quantum_2018}%
  \BibitemOpen
  \bibfield  {author} {\bibinfo {author} {\bibfnamefont {B.}~\bibnamefont
  {Shanahan}}, \bibinfo {author} {\bibfnamefont {A.}~\bibnamefont {Chenu}},
  \bibinfo {author} {\bibfnamefont {N.}~\bibnamefont {Margolus}}, \ and\
  \bibinfo {author} {\bibfnamefont {A.}~\bibnamefont {del Campo}},\ }\bibfield
  {title} {\enquote {\bibinfo {title} {Quantum {Speed} {Limits} across the
  {Quantum}-to-{Classical} {Transition}},}\ }\href {\doibase
  10.1103/PhysRevLett.120.070401} {\bibfield  {journal} {\bibinfo  {journal}
  {Physical Review Letters}\ }\textbf {\bibinfo {volume} {120}},\ \bibinfo
  {pages} {070401} (\bibinfo {year} {2018})}\BibitemShut {NoStop}%
\bibitem [{\citenamefont {Funo}, \citenamefont {Shiraishi},\ and\ \citenamefont
  {Saito}(2019)}]{funo_speed_2019}%
  \BibitemOpen
  \bibfield  {author} {\bibinfo {author} {\bibfnamefont {K.}~\bibnamefont
  {Funo}}, \bibinfo {author} {\bibfnamefont {N.}~\bibnamefont {Shiraishi}}, \
  and\ \bibinfo {author} {\bibfnamefont {K.}~\bibnamefont {Saito}},\ }\bibfield
   {title} {\enquote {\bibinfo {title} {Speed limit for open quantum
  systems},}\ }\href {\doibase 10.1088/1367-2630/aaf9f5} {\bibfield  {journal}
  {\bibinfo  {journal} {New Journal of Physics}\ }\textbf {\bibinfo {volume}
  {21}},\ \bibinfo {pages} {013006} (\bibinfo {year} {2019})}\BibitemShut
  {NoStop}%
\bibitem [{\citenamefont {Shiraishi}\ and\ \citenamefont
  {Saito}(2021)}]{shiraishi_speed_2020}%
  \BibitemOpen
  \bibfield  {author} {\bibinfo {author} {\bibfnamefont {N.}~\bibnamefont
  {Shiraishi}}\ and\ \bibinfo {author} {\bibfnamefont {K.}~\bibnamefont
  {Saito}},\ }\bibfield  {title} {\enquote {\bibinfo {title} {Speed limit for
  open systems coupled to general environments},}\ }\href {\doibase
  10.1103/PhysRevResearch.3.023074} {\bibfield  {journal} {\bibinfo  {journal}
  {Phys. Rev. Research}\ }\textbf {\bibinfo {volume} {3}},\ \bibinfo {pages}
  {023074} (\bibinfo {year} {2021})}\BibitemShut {NoStop}%
\bibitem [{\citenamefont {Plata}\ \emph {et~al.}(2020)\citenamefont {Plata},
  \citenamefont {Guéry-Odelin}, \citenamefont {Trizac},\ and\ \citenamefont
  {Prados}}]{plata_finite-time_2020}%
  \BibitemOpen
  \bibfield  {author} {\bibinfo {author} {\bibfnamefont {C.~A.}\ \bibnamefont
  {Plata}}, \bibinfo {author} {\bibfnamefont {D.}~\bibnamefont
  {Guéry-Odelin}}, \bibinfo {author} {\bibfnamefont {E.}~\bibnamefont
  {Trizac}}, \ and\ \bibinfo {author} {\bibfnamefont {A.}~\bibnamefont
  {Prados}},\ }\bibfield  {title} {\enquote {\bibinfo {title} {Finite-time
  adiabatic processes: {Derivation} and speed limit},}\ }\href {\doibase
  10.1103/PhysRevE.101.032129} {\bibfield  {journal} {\bibinfo  {journal}
  {Physical Review E}\ }\textbf {\bibinfo {volume} {101}},\ \bibinfo {pages}
  {032129} (\bibinfo {year} {2020})}\BibitemShut {NoStop}%
\bibitem [{\citenamefont {Deffner}(2020)}]{deffner_quantum_2020}%
  \BibitemOpen
  \bibfield  {author} {\bibinfo {author} {\bibfnamefont {S.}~\bibnamefont
  {Deffner}},\ }\bibfield  {title} {\enquote {\bibinfo {title} {Quantum speed
  limits and the maximal rate of information production},}\ }\href {\doibase
  10.1103/PhysRevResearch.2.013161} {\bibfield  {journal} {\bibinfo  {journal}
  {Physical Review Research}\ }\textbf {\bibinfo {volume} {2}},\ \bibinfo
  {pages} {013161} (\bibinfo {year} {2020})}\BibitemShut {NoStop}%
\bibitem [{\citenamefont {Ito}\ and\ \citenamefont
  {Dechant}(2020)}]{ito_stochastic_2020}%
  \BibitemOpen
  \bibfield  {author} {\bibinfo {author} {\bibfnamefont {S.}~\bibnamefont
  {Ito}}\ and\ \bibinfo {author} {\bibfnamefont {A.}~\bibnamefont {Dechant}},\
  }\bibfield  {title} {\enquote {\bibinfo {title} {Stochastic time-evolution,
  information geometry and the {Cramer}-{Rao} {Bound}},}\ }\href {\doibase
  10.1103/PhysRevX.10.021056} {\bibfield  {journal} {\bibinfo  {journal}
  {Physical Review X}\ }\textbf {\bibinfo {volume} {10}},\ \bibinfo {pages}
  {021056} (\bibinfo {year} {2020})}\BibitemShut {NoStop}%
\bibitem [{\citenamefont {Van~Vu}\ and\ \citenamefont
  {Hasegawa}(2021)}]{van_vu_geometrical_2021}%
  \BibitemOpen
  \bibfield  {author} {\bibinfo {author} {\bibfnamefont {T.}~\bibnamefont
  {Van~Vu}}\ and\ \bibinfo {author} {\bibfnamefont {Y.}~\bibnamefont
  {Hasegawa}},\ }\bibfield  {title} {\enquote {\bibinfo {title} {Geometrical
  {Bounds} of the {Irreversibility} in {Markovian} {Systems}},}\ }\href
  {\doibase 10.1103/PhysRevLett.126.010601} {\bibfield  {journal} {\bibinfo
  {journal} {Physical Review Letters}\ }\textbf {\bibinfo {volume} {126}},\
  \bibinfo {pages} {010601} (\bibinfo {year} {2021})}\BibitemShut {NoStop}%
\bibitem [{\citenamefont {Prados}(2021)}]{prados_optimizing_2021}%
  \BibitemOpen
  \bibfield  {author} {\bibinfo {author} {\bibfnamefont {A.}~\bibnamefont
  {Prados}},\ }\bibfield  {title} {\enquote {\bibinfo {title} {Optimizing the
  relaxation route with optimal control},}\ }\href {\doibase
  10.1103/PhysRevResearch.3.023128} {\bibfield  {journal} {\bibinfo  {journal}
  {Physical Review Research}\ }\textbf {\bibinfo {volume} {3}},\ \bibinfo
  {pages} {023128} (\bibinfo {year} {2021})}\BibitemShut {NoStop}%
\bibitem [{\citenamefont {Lee}\ \emph {et~al.}(2022)\citenamefont {Lee},
  \citenamefont {Lee}, \citenamefont {Kwon},\ and\ \citenamefont
  {Park}}]{lee_speed_2022}%
  \BibitemOpen
  \bibfield  {author} {\bibinfo {author} {\bibfnamefont {J.~S.}\ \bibnamefont
  {Lee}}, \bibinfo {author} {\bibfnamefont {S.}~\bibnamefont {Lee}}, \bibinfo
  {author} {\bibfnamefont {H.}~\bibnamefont {Kwon}}, \ and\ \bibinfo {author}
  {\bibfnamefont {H.}~\bibnamefont {Park}},\ }\bibfield  {title} {\enquote
  {\bibinfo {title} {Speed {Limit} for a {Highly} {Irreversible} {Process} and
  {Tight} {Finite}-{Time} {Landauer}’s {Bound}},}\ }\href {\doibase
  10.1103/PhysRevLett.129.120603} {\bibfield  {journal} {\bibinfo  {journal}
  {Physical Review Letters}\ }\textbf {\bibinfo {volume} {129}},\ \bibinfo
  {pages} {120603} (\bibinfo {year} {2022})}\BibitemShut {NoStop}%
\bibitem [{\citenamefont {Patrón}, \citenamefont {Prados},\ and\ \citenamefont
  {Plata}(2022)}]{patron_thermal_2022}%
  \BibitemOpen
  \bibfield  {author} {\bibinfo {author} {\bibfnamefont {A.}~\bibnamefont
  {Patrón}}, \bibinfo {author} {\bibfnamefont {A.}~\bibnamefont {Prados}}, \
  and\ \bibinfo {author} {\bibfnamefont {C.~A.}\ \bibnamefont {Plata}},\
  }\bibfield  {title} {\enquote {\bibinfo {title} {Thermal brachistochrone for
  harmonically confined {Brownian} particles},}\ }\href {\doibase
  10.1140/epjp/s13360-022-03150-3} {\bibfield  {journal} {\bibinfo  {journal}
  {The European Physical Journal Plus}\ }\textbf {\bibinfo {volume} {137}},\
  \bibinfo {pages} {1011} (\bibinfo {year} {2022})}\BibitemShut {NoStop}%
\bibitem [{\citenamefont {Dechant}(2022)}]{dechant_minimum_2022}%
  \BibitemOpen
  \bibfield  {author} {\bibinfo {author} {\bibfnamefont {A.}~\bibnamefont
  {Dechant}},\ }\bibfield  {title} {\enquote {\bibinfo {title} {Minimum entropy
  production, detailed balance and {Wasserstein} distance for continuous-time
  {Markov} processes},}\ }\href {\doibase 10.1088/1751-8121/ac4ac0} {\bibfield
  {journal} {\bibinfo  {journal} {Journal of Physics A: Mathematical and
  Theoretical}\ }\textbf {\bibinfo {volume} {55}},\ \bibinfo {pages} {094001}
  (\bibinfo {year} {2022})}\BibitemShut {NoStop}%
\bibitem [{\citenamefont {Gelfand}\ and\ \citenamefont
  {Fomin}(2000)}]{gelfand_calculus_2000}%
  \BibitemOpen
  \bibfield  {author} {\bibinfo {author} {\bibfnamefont {I.~M.}\ \bibnamefont
  {Gelfand}}\ and\ \bibinfo {author} {\bibfnamefont {S.~V.}\ \bibnamefont
  {Fomin}},\ }\href@noop {} {\emph {\bibinfo {title} {Calculus of
  {Variations}}}}\ (\bibinfo  {publisher} {Dover Publications},\ \bibinfo
  {year} {2000})\BibitemShut {NoStop}%
\bibitem [{\citenamefont {Liberzon}(2012)}]{liberzon_calculus_2012}%
  \BibitemOpen
  \bibfield  {author} {\bibinfo {author} {\bibfnamefont {D.}~\bibnamefont
  {Liberzon}},\ }\href@noop {} {\emph {\bibinfo {title} {Calculus of
  {Variations} and {Optimal} {Control} {Theory}: {A} {Concise}
  {Introduction}}}}\ (\bibinfo  {publisher} {Princeton University Press},\
  \bibinfo {year} {2012})\BibitemShut {NoStop}%
\bibitem [{Note1()}]{Note1}%
  \BibitemOpen
  \bibinfo {note} {Note that, at variance with simpler problems with linear
  damping, the sign of $W_{\protect \text {nc}}$ is not fixed because the
  factor $x_1^2-1$ changes sign at $x_1=1$.}\BibitemShut {Stop}%
\bibitem [{\citenamefont {Blaber}\ and\ \citenamefont
  {Sivak}(2023)}]{blaber_optimal_2023}%
  \BibitemOpen
  \bibfield  {author} {\bibinfo {author} {\bibfnamefont {S.}~\bibnamefont
  {Blaber}}\ and\ \bibinfo {author} {\bibfnamefont {D.~A.}\ \bibnamefont
  {Sivak}},\ }\bibfield  {title} {\enquote {\bibinfo {title} {Optimal control
  in stochastic thermodynamics},}\ }\href {\doibase 10.1088/2399-6528/acbf04}
  {\bibfield  {journal} {\bibinfo  {journal} {Journal of Physics
  Communications}\ }\textbf {\bibinfo {volume} {7}},\ \bibinfo {pages} {033001}
  (\bibinfo {year} {2023})}\BibitemShut {NoStop}%
\bibitem [{Note2()}]{Note2}%
  \BibitemOpen
  \bibinfo {note} {For fixed initial and final points of the trajectory on
  phase plane, $\Delta E$ has a given value. In that case, the problem of
  minimising the non-conservative work $W_{\protect \text {nc}}$ and the total
  work done by the external force $F(t)$ are always equivalent. See also
  Sec.~\ref {sec:min-total-work}.}\BibitemShut {Stop}%
\bibitem [{\citenamefont {Turner}, \citenamefont {McClintock},\ and\
  \citenamefont {Stefanovska}(2015)}]{turner_maximum_2015}%
  \BibitemOpen
  \bibfield  {author} {\bibinfo {author} {\bibfnamefont {N.}~\bibnamefont
  {Turner}}, \bibinfo {author} {\bibfnamefont {P.~V.~E.}\ \bibnamefont
  {McClintock}}, \ and\ \bibinfo {author} {\bibfnamefont {A.}~\bibnamefont
  {Stefanovska}},\ }\bibfield  {title} {\enquote {\bibinfo {title} {Maximum
  amplitude of limit cycles in {Liénard} systems},}\ }\href {\doibase
  10.1103/PhysRevE.91.012927} {\bibfield  {journal} {\bibinfo  {journal}
  {Physical Review E}\ }\textbf {\bibinfo {volume} {91}},\ \bibinfo {pages}
  {012927} (\bibinfo {year} {2015})}\BibitemShut {NoStop}%
\bibitem [{Note3()}]{Note3}%
  \BibitemOpen
  \bibinfo {note} {Note that this definition of $H$, which is the one employed
  in classical mechanics, leads to the usual definition in Pontryagin's theory
  of optimal control.}\BibitemShut {Stop}%
\bibitem [{\citenamefont {Robbins}(1967)}]{robbins_generalized_1967}%
  \BibitemOpen
  \bibfield  {author} {\bibinfo {author} {\bibfnamefont {H.~M.}\ \bibnamefont
  {Robbins}},\ }\bibfield  {title} {\enquote {\bibinfo {title} {A {Generalized}
  {Legendre}-{Clebsch} {Condition} for the {Singular} {Cases} of {Optimal}
  {Control}},}\ }\href {\doibase 10.1147/rd.114.0361} {\bibfield  {journal}
  {\bibinfo  {journal} {IBM Journal of Research and Development}\ }\textbf
  {\bibinfo {volume} {11}},\ \bibinfo {pages} {361--372} (\bibinfo {year}
  {1967})}\BibitemShut {NoStop}%
\bibitem [{\citenamefont {Gomez-Marin}, \citenamefont {Schmiedl},\ and\
  \citenamefont {Seifert}(2008)}]{gomez-marin_optimal_2008}%
  \BibitemOpen
  \bibfield  {author} {\bibinfo {author} {\bibfnamefont {A.}~\bibnamefont
  {Gomez-Marin}}, \bibinfo {author} {\bibfnamefont {T.}~\bibnamefont
  {Schmiedl}}, \ and\ \bibinfo {author} {\bibfnamefont {U.}~\bibnamefont
  {Seifert}},\ }\bibfield  {title} {\enquote {\bibinfo {title} {Optimal
  protocols for minimal work processes in underdamped stochastic
  thermodynamics},}\ }\href {\doibase 10.1063/1.2948948} {\bibfield  {journal}
  {\bibinfo  {journal} {The Journal of Chemical Physics}\ }\textbf {\bibinfo
  {volume} {129}},\ \bibinfo {pages} {024114} (\bibinfo {year}
  {2008})}\BibitemShut {NoStop}%
\bibitem [{Note4()}]{Note4}%
  \BibitemOpen
  \bibinfo {note} {Due to the symmetry $x_1\to -x_1$ of our problem, all the
  plots in this section are for $|x_{10}|\ge 1$.}\BibitemShut {Stop}%
\bibitem [{Note5()}]{Note5}%
  \BibitemOpen
  \bibinfo {note} {For an arbitrary value of $\mu $, the proportionality
  constant also depends { very weakly} on the damping coefficient $\mu $
  through $x_{\ell c}^{\protect \qopname \relax m{max}}$.}\BibitemShut {Stop}%
\bibitem [{\citenamefont {Plata}\ \emph {et~al.}(2021)\citenamefont {Plata},
  \citenamefont {Prados}, \citenamefont {Trizac},\ and\ \citenamefont
  {Gu{\'e}ry-Odelin}}]{plata_taming_2021}%
  \BibitemOpen
  \bibfield  {author} {\bibinfo {author} {\bibfnamefont {C.~A.}\ \bibnamefont
  {Plata}}, \bibinfo {author} {\bibfnamefont {A.}~\bibnamefont {Prados}},
  \bibinfo {author} {\bibfnamefont {E.}~\bibnamefont {Trizac}}, \ and\ \bibinfo
  {author} {\bibfnamefont {D.}~\bibnamefont {Gu{\'e}ry-Odelin}},\ }\bibfield
  {title} {\enquote {\bibinfo {title} {Taming the {Time} {Evolution} in
  {Overdamped} {Systems}: {Shortcuts} {Elaborated} from {Fast}-{Forward} and
  {Time}-{Reversed} {Protocols}},}\ }\href {\doibase
  10.1103/PhysRevLett.127.190605} {\bibfield  {journal} {\bibinfo  {journal}
  {Physical Review Letters}\ }\textbf {\bibinfo {volume} {127}},\ \bibinfo
  {pages} {190605} (\bibinfo {year} {2021})}\BibitemShut {NoStop}%
\bibitem [{Note6()}]{Note6}%
  \BibitemOpen
  \bibinfo {note} {For $\mu =0.1$, $x_{\ell c}^{\protect \qopname \relax
  m{max}}=2.00010$.\cite {turner_maximum_2015}}\BibitemShut {NoStop}%
\bibitem [{\citenamefont {Iacono}\ and\ \citenamefont
  {Russo}(2011)}]{iacono_class_2011}%
  \BibitemOpen
  \bibfield  {author} {\bibinfo {author} {\bibfnamefont {R.}~\bibnamefont
  {Iacono}}\ and\ \bibinfo {author} {\bibfnamefont {F.}~\bibnamefont {Russo}},\
  }\bibfield  {title} {\enquote {\bibinfo {title} {Class of solvable nonlinear
  oscillators with isochronous orbits},}\ }\href {\doibase
  10.1103/PhysRevE.83.027601} {\bibfield  {journal} {\bibinfo  {journal}
  {Physical Review E}\ }\textbf {\bibinfo {volume} {83}},\ \bibinfo {pages}
  {027601} (\bibinfo {year} {2011})}\BibitemShut {NoStop}%
\bibitem [{\citenamefont {Messias}\ and\ \citenamefont
  {Alves~Gouveia}(2011)}]{messias_time-periodic_2011}%
  \BibitemOpen
  \bibfield  {author} {\bibinfo {author} {\bibfnamefont {M.}~\bibnamefont
  {Messias}}\ and\ \bibinfo {author} {\bibfnamefont {M.~R.}\ \bibnamefont
  {Alves~Gouveia}},\ }\bibfield  {title} {\enquote {\bibinfo {title}
  {Time-periodic perturbation of a {Liénard} equation with an unbounded
  homoclinic loop},}\ }\href {\doibase 10.1016/j.physd.2011.06.006} {\bibfield
  {journal} {\bibinfo  {journal} {Physica D: Nonlinear Phenomena}\ }\textbf
  {\bibinfo {volume} {240}},\ \bibinfo {pages} {1402--1409} (\bibinfo {year}
  {2011})}\BibitemShut {NoStop}%
\bibitem [{\citenamefont {Ghosh}\ and\ \citenamefont
  {Ray}(2014)}]{ghosh_lienard-type_2014}%
  \BibitemOpen
  \bibfield  {author} {\bibinfo {author} {\bibfnamefont {S.}~\bibnamefont
  {Ghosh}}\ and\ \bibinfo {author} {\bibfnamefont {D.~S.}\ \bibnamefont
  {Ray}},\ }\bibfield  {title} {\enquote {\bibinfo {title} {Liénard-type
  chemical oscillator},}\ }\href {\doibase 10.1140/epjb/e2014-41070-1}
  {\bibfield  {journal} {\bibinfo  {journal} {The European Physical Journal B}\
  }\textbf {\bibinfo {volume} {87}},\ \bibinfo {pages} {65} (\bibinfo {year}
  {2014})}\BibitemShut {NoStop}%
\bibitem [{\citenamefont {Shah}\ \emph {et~al.}(2015)\citenamefont {Shah},
  \citenamefont {Chattopadhyay}, \citenamefont {Vaidya},\ and\ \citenamefont
  {Chakraborty}}]{shah_conservative_2015}%
  \BibitemOpen
  \bibfield  {author} {\bibinfo {author} {\bibfnamefont {T.}~\bibnamefont
  {Shah}}, \bibinfo {author} {\bibfnamefont {R.}~\bibnamefont {Chattopadhyay}},
  \bibinfo {author} {\bibfnamefont {K.}~\bibnamefont {Vaidya}}, \ and\ \bibinfo
  {author} {\bibfnamefont {S.}~\bibnamefont {Chakraborty}},\ }\bibfield
  {title} {\enquote {\bibinfo {title} {Conservative perturbation theory for
  nonconservative systems},}\ }\href {\doibase 10.1103/PhysRevE.92.062927}
  {\bibfield  {journal} {\bibinfo  {journal} {Physical Review E}\ }\textbf
  {\bibinfo {volume} {92}},\ \bibinfo {pages} {062927} (\bibinfo {year}
  {2015})}\BibitemShut {NoStop}%
\bibitem [{\citenamefont {Giné}(2017)}]{gine_lienard_2017}%
  \BibitemOpen
  \bibfield  {author} {\bibinfo {author} {\bibfnamefont {J.}~\bibnamefont
  {Giné}},\ }\bibfield  {title} {\enquote {\bibinfo {title} {Liénard
  {Equation} and {Its} {Generalizations}},}\ }\href {\doibase
  10.1142/S021812741750081X} {\bibfield  {journal} {\bibinfo  {journal}
  {International Journal of Bifurcation and Chaos}\ }\textbf {\bibinfo {volume}
  {27}},\ \bibinfo {pages} {1750081} (\bibinfo {year} {2017})}\BibitemShut
  {NoStop}%
\bibitem [{\citenamefont {Mishra}, \citenamefont {Saha},\ and\ \citenamefont
  {Dana}(2023)}]{mishra_chimeras_2023}%
  \BibitemOpen
  \bibfield  {author} {\bibinfo {author} {\bibfnamefont {A.}~\bibnamefont
  {Mishra}}, \bibinfo {author} {\bibfnamefont {S.}~\bibnamefont {Saha}}, \ and\
  \bibinfo {author} {\bibfnamefont {S.~K.}\ \bibnamefont {Dana}},\ }\bibfield
  {title} {\enquote {\bibinfo {title} {Chimeras in globally coupled
  oscillators: {A} review},}\ }\href {\doibase 10.1063/5.0143872} {\bibfield
  {journal} {\bibinfo  {journal} {Chaos: An Interdisciplinary Journal of
  Nonlinear Science}\ }\textbf {\bibinfo {volume} {33}},\ \bibinfo {pages}
  {092101} (\bibinfo {year} {2023})}\BibitemShut {NoStop}%
\bibitem [{Note7()}]{Note7}%
  \BibitemOpen
  \bibinfo {note} {For the van der Pol equation, $\xi (x)=(x^3-3x)/3$ and
  $a=\protect \sqrt {3}$.}\BibitemShut {Stop}%
\bibitem [{Note8()}]{Note8}%
  \BibitemOpen
  \bibinfo {note} {Similarly to the discussion below Eq.~\protect \textup
  {\hbox {\mathsurround \z@ \protect \normalfont (\ignorespaces \ref
  {eq:p1tf-trans}\unskip \@@italiccorr )}}, the case $h(x_1(t_f^-))=0$ is a
  particular case of $x_2(t_f^-)=0$.}\BibitemShut {Stop}%
\bibitem [{\citenamefont {Hu}, \citenamefont {Li},\ and\ \citenamefont
  {Lv}(2021)}]{hu_phototunable_2021}%
  \BibitemOpen
  \bibfield  {author} {\bibinfo {author} {\bibfnamefont {Z.}~\bibnamefont
  {Hu}}, \bibinfo {author} {\bibfnamefont {Y.}~\bibnamefont {Li}}, \ and\
  \bibinfo {author} {\bibfnamefont {J.-a.}\ \bibnamefont {Lv}},\ }\bibfield
  {title} {\enquote {\bibinfo {title} {Phototunable self-oscillating system
  driven by a self-winding fiber actuator},}\ }\href {\doibase
  10.1038/s41467-021-23562-6} {\bibfield  {journal} {\bibinfo  {journal}
  {Nature Communications}\ }\textbf {\bibinfo {volume} {12}},\ \bibinfo {pages}
  {3211} (\bibinfo {year} {2021})}\BibitemShut {NoStop}%
\bibitem [{\citenamefont {Landman}, \citenamefont {Verduyn~Lunel},\ and\
  \citenamefont {Kegel}(2023)}]{landman_transcription_2023}%
  \BibitemOpen
  \bibfield  {author} {\bibinfo {author} {\bibfnamefont {J.}~\bibnamefont
  {Landman}}, \bibinfo {author} {\bibfnamefont {S.~M.}\ \bibnamefont
  {Verduyn~Lunel}}, \ and\ \bibinfo {author} {\bibfnamefont {W.~K.}\
  \bibnamefont {Kegel}},\ }\bibfield  {title} {\enquote {\bibinfo {title}
  {Transcription factor competition facilitates self-sustained oscillations in
  single gene genetic circuits},}\ }\href {\doibase
  10.1371/journal.pcbi.1011525} {\bibfield  {journal} {\bibinfo  {journal}
  {PLOS Computational Biology}\ }\textbf {\bibinfo {volume} {19}},\ \bibinfo
  {pages} {e1011525} (\bibinfo {year} {2023})}\BibitemShut {NoStop}%
\bibitem [{\citenamefont {Zheng}\ \emph {et~al.}(2023)\citenamefont {Zheng},
  \citenamefont {Brandenbourger}, \citenamefont {Robinet}, \citenamefont
  {Schall}, \citenamefont {Lerner},\ and\ \citenamefont
  {Coulais}}]{zheng_self-oscillation_2023}%
  \BibitemOpen
  \bibfield  {author} {\bibinfo {author} {\bibfnamefont {E.}~\bibnamefont
  {Zheng}}, \bibinfo {author} {\bibfnamefont {M.}~\bibnamefont
  {Brandenbourger}}, \bibinfo {author} {\bibfnamefont {L.}~\bibnamefont
  {Robinet}}, \bibinfo {author} {\bibfnamefont {P.}~\bibnamefont {Schall}},
  \bibinfo {author} {\bibfnamefont {E.}~\bibnamefont {Lerner}}, \ and\ \bibinfo
  {author} {\bibfnamefont {C.}~\bibnamefont {Coulais}},\ }\bibfield  {title}
  {\enquote {\bibinfo {title} {Self-{Oscillation} and {Synchronization}
  {Transitions} in {Elastoactive} {Structures}},}\ }\href {\doibase
  10.1103/PhysRevLett.130.178202} {\bibfield  {journal} {\bibinfo  {journal}
  {Physical Review Letters}\ }\textbf {\bibinfo {volume} {130}},\ \bibinfo
  {pages} {178202} (\bibinfo {year} {2023})}\BibitemShut {NoStop}%
\bibitem [{\citenamefont {Pontryagin}(1987)}]{pontryagin_mathematical_1987}%
  \BibitemOpen
  \bibfield  {author} {\bibinfo {author} {\bibfnamefont {L.~S.}\ \bibnamefont
  {Pontryagin}},\ }\href@noop {} {\emph {\bibinfo {title} {Mathematical
  {Theory} of {Optimal} {Processes}}}}\ (\bibinfo  {publisher} {CRC Press},\
  \bibinfo {year} {1987})\BibitemShut {NoStop}%
\bibitem [{Note9()}]{Note9}%
  \BibitemOpen
  \bibinfo {note} {In fact, this is the case for other underdamped control
  problems, see for instance Ref.~\cite {baldovin_levitated_2024}.}\BibitemShut
  {Stop}%
\bibitem [{\citenamefont {Landau}\ and\ \citenamefont
  {Lifshitz}(1976)}]{landau_mechanics_1976}%
  \BibitemOpen
  \bibfield  {author} {\bibinfo {author} {\bibfnamefont {L.~D.}\ \bibnamefont
  {Landau}}\ and\ \bibinfo {author} {\bibfnamefont {E.~M.}\ \bibnamefont
  {Lifshitz}},\ }\href@noop {} {\emph {\bibinfo {title} {Course of Theoretical
  Physics: Vol. 1, Mechanics}}},\ \bibinfo {edition} {3rd}\ ed.,\ Course of
  Theoretical Physics\ (\bibinfo  {publisher} {Butterworth-Heinemann},\
  \bibinfo {year} {1976})\BibitemShut {NoStop}%
\bibitem [{\citenamefont {Baldovin}\ \emph {et~al.}(2024)\citenamefont
  {Baldovin}, \citenamefont {Yedder}, \citenamefont {Plata}, \citenamefont
  {Raynal}, \citenamefont {Rondin}, \citenamefont {Trizac},\ and\ \citenamefont
  {Prados}}]{baldovin_levitated_2024}%
  \BibitemOpen
  \bibfield  {author} {\bibinfo {author} {\bibfnamefont {M.}~\bibnamefont
  {Baldovin}}, \bibinfo {author} {\bibfnamefont {I.~B.}\ \bibnamefont
  {Yedder}}, \bibinfo {author} {\bibfnamefont {C.~A.}\ \bibnamefont {Plata}},
  \bibinfo {author} {\bibfnamefont {D.}~\bibnamefont {Raynal}}, \bibinfo
  {author} {\bibfnamefont {L.}~\bibnamefont {Rondin}}, \bibinfo {author}
  {\bibfnamefont {E.}~\bibnamefont {Trizac}}, \ and\ \bibinfo {author}
  {\bibfnamefont {A.}~\bibnamefont {Prados}},\ }\href
  {https://arxiv.org/abs/2408.00043} {\enquote {\bibinfo {title} {Optimal
  control of levitated nanoparticles through finite-stiffness confinement},}\ }
  (\bibinfo {year} {2024}),\ \Eprint {http://arxiv.org/abs/2408.00043}
  {arXiv:2408.00043 [cond-mat.stat-mech]} \BibitemShut {NoStop}%
\end{thebibliography}%

\end{document}